\documentclass[journal]{IEEEtran}

\usepackage{cite}
\usepackage{balance}

\ifCLASSINFOpdf
\usepackage[pdftex]{graphicx}
\usepackage{epstopdf}
\usepackage{subcaption}
\usepackage{pgfplots}
\usepackage{amsmath,stmaryrd,pict2e,picture}
\else
\fi
\usepackage{amsmath,amssymb,amsthm}
\usepackage{mathrsfs}
\usepackage{algorithm, algorithmic}

    \usepackage{bm}
\usepackage{accents}
\newlength{\dhatheight}
\newcommand{\doublehat}[1]{%
	\settoheight{\dhatheight}{\ensuremath{\hat{#1}}}%
	\addtolength{\dhatheight}{-0.35ex}%
	\hat{\vphantom{\rule{1pt}{\dhatheight}}%
		\smash{\hat{#1}}}}

\begin{document}
	%
	\title{Sparse Array  Beamforming Design for Wideband Signal Models}
	%
	%
	%
	
	\author{Syed~A.~Hamza,~\IEEEmembership{Member,~IEEE}
		and~Moeness~G.~Amin,~\IEEEmembership{Fellow,~IEEE}
		\thanks{This work was supported by National Science Foundation (NSF) award AST-1547420. This paper was presented in part at the   52nd  Asilomar  Conference  on  Signals,  Systems,  and Computers, Oct 2018 \cite{8645552}  and in part at the IEEE Radar Conference, Boston, MA, April 2019 \cite{Hamza_2019wb}. (Corresponding author: Syed A. Hamza)
		
Syed A. Hamza is with the School of Engineering, Widener University PA and Moeness Amin is with the Center for Advanced Communications, College of Engineering, Villanova University, Villanova, PA 19085-1681 USA (e-mail: shamza@widener.edu; moeness.amin@villanova.edu).}
	}

\maketitle
\begin{abstract}
We develop   sparse array receive  beamformer design methods achieving   maximum signal-to-interference plus  noise ratio (MaxSINR)  for wideband sources and jammers. Both tapped delay line (TDL) filtering and the DFT realizations to wideband array processing are considered. The array sparsity stems from  the limited number of available RF transmission chains that switch between the sensors, thereby configuring different arrays at different times.  The sparse array configuration design problem is formulated   as a quadratically constraint quadratic program (QCQP) and solved by using SDR (semidefinite relaxation). A   computationally viable approach through SCA (successive convex relaxation) is also pursued.  In order to realize an implementable design, in presence of missing autocorrelation lags, we propose  parameter-free block Toeplitz matrix completion to estimate the received data correlation matrix across the entire array aperture.   It is shown that the optimum wideband sparse array effectively utilizes the array aperture  and provides  considerable performance improvement over  suboptimal array topologies. 
\end{abstract}
\IEEEpeerreviewmaketitle
\section{Introduction}  \label{Introduction}
%
%
%
%

Wideband systems  can deliver accurate target localization   for radar systems \cite{1359140}, provide diversity, reliability and anti jamming capabilities to the wireless communication systems and signal enhancement for microphone arrays \cite{Brandstein2001MicrophoneA, Goldsmith:2005:WC:993515},   whereas the  UWB (Ultra-wideband) systems play a major role in high resolution imagery in medical imaging \cite{2f1ce6c9e62d4b60acddd36683f19924, 5666802}.  

\begin{figure}[!t]
	\centering
	\includegraphics[width=3.28 in, height=1.75 in]{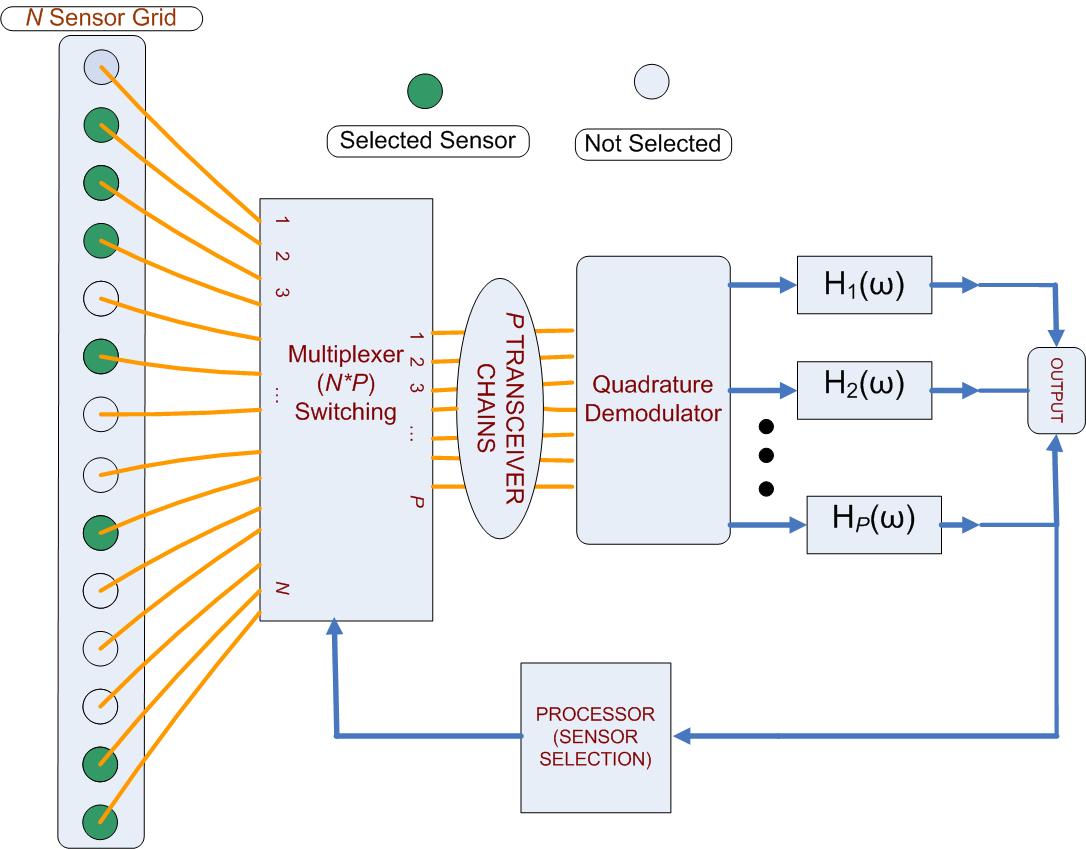}
	\caption{Block Diagram of sparse array wideband processing.}
	\label{labelfig1}
\end{figure}

 Beamforming techniques for wideband signals  either involve  a fixed design such as a frequency invariant beamformer  or an adaptive design  based on  the linearly constrained minimum variance (LCMV) beamformer \cite{doi:10.1002/0471733482.ch7, Liu2010WidebandBC}.  Irrespective of the design criterion, the wideband beamformers  are typically  implemented jointly in the spatial and temporal domains, as shown in  Fig. \ref{labelfig1}. This spatio-temporal processing is often realized through   two different  schemes, namely, the tapped delay line  (TDL) filtering  or subband processing like  DFT. For the former, an $L$ TDL filter is assumed for each sensor, and the received data at each sampling instant is processed for all sensors jointly \cite{1165142, 665}.  In the DFT implementation scheme, the data at each sensor is buffered and transformed to the frequency domain by $L$-point DFT. Once expressed in terms of narrowband signals,  optimal beamforming is performed in each DFT bin.  The DFT implementation is computationally more viable \cite{1450747, 1164219, 1143832}. However, the TDL   implementation scheme has an added advantage since buffering is not required and  the spatio-temporal  weight vector can be updated at each sampling instant.   The TDL and DFT beamformers would render  identical output signal  if the corresponding beamformer weights are   related through  a DFT transformation. However, carrying  the beamformer design,  separately in each domain, doesn't  warrant  the beamformer weights forming strictly a  DFT pair.  Resultantly, the output could  differ  slightly for each implementation. To circumvent the  computationally expensive  TDL beamformer design, a DFT beamformer is rather optimized and  the DFT transformation is subsequently used  to obtain  the TDL beamformer. This dual domain  TDL implementation designed primarily in the DFT domain   can  yield adequate output performance in practice \cite{ doi:10.1121/1.413765}.

 Sparse array design strives to optimally deploy  sensors, essentially   achieving desirable beamforming characteristics, lowering  the system hardware costs and  reducing the computational  complexity. Sparse array design is known to yield considerable performance advantages under different design criteria for narrowband signal models.  These   criteria can largely be segregated  into environment-independent  and environment-dependent designs. The minimum redundancy arrays (MRA) and the structured sparse array configurations are cases of the former design. They seek to optimize the environment-blind design criterion to enable the DOA estimation of more sources than physical sensors \cite{5739227,  7012090, 5456168, 7106511}.  More recently, the   switched antenna and beam technologies  have motivated  the design for environment  adaptive sparse arrays. The available sensors are switched  according to  changing environmental conditions, enabling optimum use of expensive transceiver chains  \cite{738234, 1299086, 922993, 967086}. Sparse array design based on Cramer-Rao lower bound (CRLB) minimization criterion was shown effective  for DOA estimation schemes \cite{8386702}.   On the other hand,   beamforming approaches typically implements MaxSINR criterion, yielding efficient  adaptive  performance  that is dependent on the underlying operating environment \cite{6774934, 8036237, 8061016,  663844798989898, 6477161, 1634819, 5680595}. 

 Designing sparse arrays has proved to be particularly advantageous for wideband signal models, as it circumvents the contradicting design requirements posed by the frequency spread of the signal. On the one hand,  the lower frequency end of the spectrum puts a minimum aperture constraint on the array to achieve  certain resolution. On the other hand, the higher end of the spectrum dictates  the minimum separation between consecutive sensor locations so as to avoid spatial aliasing, consequently, resulting in the oversampling of the lower frequency content.   For a limited number of available sensors, wideband sparse array design can, in essence,  yield improved performance in many design applications  by offering more control over beampattern characteristics for all frequencies of interest  \cite{81155, 23483091a3104a7f8bd7d213f8ada7cc, 1071, MURINO1997177}.  Many different  metrics,  such as frequency invariant beampattern synthesis, robust beampattern design   and side-lobe level control   have been proposed for optimal  wideband sparse array beamforming \cite{482017,  doi:10.1121/1.412215, 1071, 968251,  93229}.  For instance,  simulated annealing has been applied in \cite{968251} to achieve the desired peak sidelobe levels, while jointly optimizing  the beamformer weights and sensor locations.   Frequency invariant beampattern optimization  for  wideband sparse array design employing  compressive sensing (CS) technique has  been implemented  in \cite{6822510}.  The authors therein,   invoked sparsity in the temporal and spatial domains, simultaneously,  in an  attempt to decrease the overall processing complexity.  
 
  In this paper, we examine the Capon based MaxSINR sparse array design from both the DFT  and TDL filtering realization perspectives.   We consider a switched array  adaptive beamformer design  which is   fundamentally different  as compared to the aforementioned wideband  performance metrics that optimize a prefixed receive beampattern for certain sector/frequencies of interest, independent of the received data statistics.      
We examine  environment-dependent sparse arrays that maximize the SINR for  frequency spread  source operating in  wideband jamming environments. The objective of the populated and the sparse wideband beamformers is fundamentally the same;  minimization of noise and interference signals at the array output while simultaneously maintaining a desired response in the direction of interest. We  adopt a Capon based methodology for  enhancing the   signal power for the desired source operating  in an interference active environment. Capon method is a well known linear constraint beamforming approach that rejects the interference and maximizes the output SINR \cite{4101326}. It  provides superior estimation accuracy but  assumes the exact knowledge  or estimated version of the received data correlation matrix across the entire array aperture. The latter is the case in many  key applications in  radar   signal processing and   medical imaging  \cite{Trees:1992:DEM:530789, 1206680, Hamza_2019wb, 8645552}. This assumption, however, cannot be readily made for sparse array configurations.
We pose the design  problem as optimally selecting $P$ sensors out of $N$ possible equally spaced locations. For the scope of this paper, we ignore any mutual coupling or sensor failure possibilities that could arise due to the closely spaced perspective locations on the grid. Each sensor has an associated  $L$ TDL or $L$ point DFT filtering  to jointly process the  signal in the temporal and spatial  domains. Our approach is a  natural extension of Capon beamforming  at the receiver  and amounts to maximizing the SINR over all possible sparse array configurations. In the case of the TDL realization, we select those sensors that  maximize the principal eigenvalue of the product of  the inverse of received data correlation matrix and the desired source  correlation matrix \cite{1223538}. For the DFT implementation scheme, the  maximization is performed over all DFT bins.  In either case, it   is an NP hard optimization problem. In order to realize convex relaxation  and avoid the computational burden of applying the singular value decomposition (SVD) for each possible configuration, we solve the underlying problem using constrained minimization based approach. We consider two different optimization approaches, namely, the semidefinite relaxation (SDR) and successive convex approximation (SCA). For SDR-based approach, we pose the problem as quadratically constraint quadratic program (QCQP) with weighted $l_{1-\infty}$-norm squared to promote group sparsity. An eigenvector based  iterative methodology is adopted  to promote group sparsity and ensure that only $P$  sensors are finally selected. It is noted that the re-weighted $l_{1-\infty}$-norm squared relaxation is shown to be  effective for reducing the number of sensors in multicast transmit beamforming \cite{6477161}. However, owing to the  computational complexity associated with the SDR approach, we alternatively  pose the  problem as  successive convex relaxation (SCA)   that approximates the problem iteratively by first order gradient approximation \cite{Ibrahim2018MirrorProxSA}. The proposed algorithms are  suitable for moderate-size antenna systems to enable real time  applications wherein the environment largely stays stationary relative to the time required for sparse configurability. 

In order to enable a data-dependent design for sparse array wideband beamforming, we require     knowledge of the   received data correlation matrix corresponding to the full array aperture. With only few active sensors at any time instant, it is  infeasible to assume such knowledge due to  missing correlation entries. We circumvent this problem by employing a low rank block Toeplitz matrix completion scheme to interpolate the missing correlation entries. Subsequently,  the interpolated data correlation matrix is input to the proposed sparse optimization algorithms.  We  demonstrate the offerings of the proposed sparse array design utilizing matrix completion under limited data snapshots by comparing its performance with that achieved through enumeration. \par

The rest of the paper is organized as follows:  In the next section, we  state the problem formulation for  maximizing the  output SINR under wideband source signal model by elucidating the TDL and DFT signal model. Section \ref{Optimum sparse array design} deals with the optimum sparse array design by  semidefinite relaxation as well as successive convex relaxation to obtain the optimum $P$  sparse array geometry. Section \ref{sec4} discusses the block Toeplitz matrix completion approach for a conceivable sparse array design from an implementation perspective. Design examples  and conclusion follow at the end.
\section{Problem Formulation} \label{Problem Formulation}
Consider  a single desired source and $Q$ interfering source signals impinging on a  linear array with $N$ uniformly placed sensors. The   Nyquist sampled received baseband signal $\mathbf{x}(n) \in \mathbb{C}^{N}$   at time instant $n$ is, therefore,  given by, 
\begin{equation} \label{a}
\mathbf{x}(n)=    \mathbf{s}(n)  + \sum_{k=1}^{Q}  \mathbf{i}_k(n) + \mathbf{v}(n),  
\end{equation}
where $\mathbf{s}(n)\in \mathbb{C}^{N} $ is the contribution from the desired signal located at $\theta_s$, $\mathbf{i}_k(n) $  is the $k$th interfering signal vectors  corresponding to the  respective direction  of arrival, $\theta_{k}$, and $\mathbf{v}(n)$ is the spatially uncorrelated sensor array output noise.
\begin{figure}[!t]
	\centering
	\includegraphics[width=3.28 in, height=1.45 in]{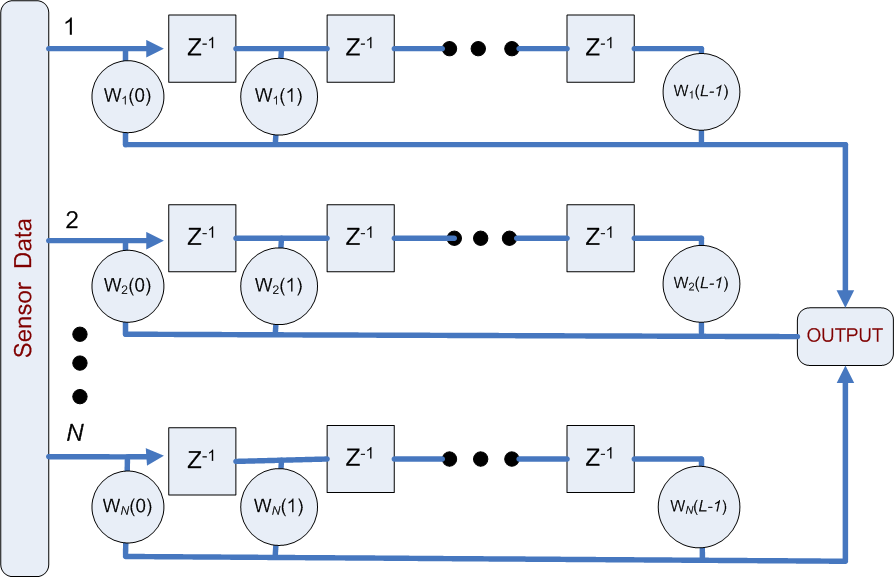}
	\caption{TDL realization of  wideband  beamforming.}
	\label{labelfig2}
\end{figure}
\subsection{TDL Implementation scheme}
We assume a  TDL of length $L$ associated with each  sensor, as shown in Fig. \ref{labelfig2}. The symbol z$^{-1}$ denotes the time delay and w$_k(m)$ is the  beamforming weight for the k$th$ sensor at m$th$ sampling instant.   We define a stacked vector $\mathbf{X}=[\mathbf{x}^T(n), \mathbf{x}^T(n-1), ..., \mathbf{x}^T(n-L+1)]^T \in \mathbb{C}^{NL}$ containing the  array data collected  over $L$ sampling instances ((.)$^T$ denotes the transpose).  Rewriting  (\ref{a}) in a compact way in terms of stacked vectors, we obtain,
\begin{equation} \label{b}
\mathbf{X}=    \mathbf{S}  + \sum_{k=1}^{Q}  \mathbf{I}_k + \mathbf{V}  
\end{equation}
Here, $\mathbf{S}=[\mathbf{s}^T(n), \mathbf{s}^T(n-1), ..., \mathbf{s}^T(n-L+1)]^T $ and similarly $\mathbf{I}_k$ and  $\mathbf{V}$ are defined, respectively, as interference and noise stacked vectors. The received signal $\mathbf{X}$  is  then  combined linearly  to maximize the output SINR. The output signal $y(n)$ of the optimum beamformer for maximum SINR is given by \cite{1223538}, 
\begin{equation}  \label{c}
y(n) = \mathbf{w}_o^H \mathbf{X},
\end{equation}
where $\mathbf{w}_o$ is the solution of the following optimization problem,
\begin{equation} \label{d}
\begin{aligned}
\underset{\mathbf{w} \in \mathbb{C}^{NL}}{\text{minimize}} & \quad   \mathbf{ w}^H\mathbf{R}_{n}\mathbf{w}\\
\text{s.t.} & \quad     \mathbf{w}^H\mathbf{R}_{s}\mathbf{ w}=1
\end{aligned}
\end{equation}
The w$_k(m)$ beamforming weight (shown in Fig. 2) maps to the $(N(m)+k)$th element of the   stacked beamformer $\mathbf{w}$ ($m \in 0, 1, ..., L-1$, $k \in 1, 2, ..., N$). Here, (.)$^H$ denotes Hermitian transpose, $\mathbf{R}_s=E\mathbf{(SS}^H) \in \mathbb{C}^{NL\times NL}$ is the desired signal correlation matrix. Likewise, $\mathbf{R}_{n}$ is the correlation matrix associated with interference and noise stacked vectors. In the case of spatial spread or wideband source signal, the correlation matrix is given by \cite{1165142},

\begin{equation}  \label{e}
 \mathbf{R}_s =  \int_{B_s}\int_{\Theta_s} \sigma^2_{\theta_s}(\omega)\mathbf{a}(\theta_s,\omega)\mathbf{a}^H(\theta_s,\omega)  d\theta_s d\omega
\end{equation} 
Here, $\sigma^2_{\theta_s}(\omega)$ is the signal power as a function of $\theta_s$ and $\omega$, $\Theta_s$ and $B_s$ are the spatial and spectral supports  of the desired source signal.
We only consider  point sources with no significant  spatial extent, hence rewriting (\ref{e}) as follows, 
\begin{equation}  \label{f}
\mathbf{R}_s =  \int_{B_s} \sigma^2_{\theta_s}(\omega)\mathbf{a}(\theta_s,\omega)\mathbf{a}^H(\theta_s,\omega) d\omega
\end{equation}
The space-time steering vector  $\mathbf{a}(\theta_s,\omega) \in \mathbb{C}^{NL}$, corresponding to the  source signal, can be represented as  a Kronecker product ($\otimes$),
\begin{equation}  \label{g}
\mathbf{a(\theta_s,\omega)} = \pmb{\phi}_{\omega}\otimes \mathbf{a}_{\theta_s}(\omega),
\end{equation}
with,
\begin{equation}  \label{h}
\pmb{\phi}_{\omega}=[1 \,  \,  \, e^{j (\pi \omega/ \omega_{max})} \,  . \,   . \,  . \, e^{j (\pi \omega/ \omega_{max}) (L-1)}]^T,
\end{equation}
\begin{equation} 
\begin{aligned}  \label{i}
\mathbf{a}_{\theta_s}(w)  &=[1 \,  \,  \, e^{j (2 \pi / \lambda_\omega) d cos(\theta_s)} \,  . \,   . \,  . \, e^{j (2 \pi / \lambda_\omega) d (N-1) cos(\theta_s)  }]^T\\
 &=[1 \,  \,  \, e^{j \pi( \frac{\omega_c+\omega}{\Omega_{max}} ) cos(\theta_s)} \,  . \,   . \,  . \, e^{j \pi( \frac{\omega_c+\omega}{\Omega_{max}} ) (N-1)cos(\theta_s)}]^T,
\end{aligned}
\end{equation}
where $\lambda_\omega$ is the wavelength corresponding to  $\omega_c+\omega$, $\omega$ and $\omega_{c}$ represent the   baseband source angular frequency and the carrier angular frequency respectively, and $\omega_{max}$ is  the maximum source baseband  angular frequency. The data is sampled temporally  at the Nyquist rate for a given signal bandwidth. Similarly, we set the inter-element spacing   $d=\lambda_{min}/2$  to avoid spatial aliasing corresponding to the highest spatial angular frequency $\Omega_{max}=\omega_c+\omega_{max}$,   where $\lambda_{min}$ is the wavelength corresponding to  $\Omega_{max}$. The correlation matrix $\mathbf{R}_k \in \mathbb{C}^{NL\times NL}$ for the   interferer $\mathbf{i}_k$  is defined according to  (\ref{f}) with respective to $\theta_k$ and  $B_k$. The sensor noise  correlation matrix, $\mathbf{R}_v=\sigma_v^2\mathbf{I} \in \mathbb{C}^{NL\times NL}$ assumes spatially and  temporally uncorrelated noise $\mathbf{v}(n)$ with variance $\sigma_v^2$.
The problem in  (\ref{d}) can  be written equivalently by replacing $\mathbf{R}_{n}=\sum_{k=1}^{Q}\mathbf{R}_k+\mathbf{R}_v$ with $\mathbf{R}=\mathbf{R}_s+ \mathbf{R}_{n}$ as follows \cite{1223538},
\begin{figure}[!t]
	\centering
	\includegraphics[width=3.28 in, height=1.45 in]{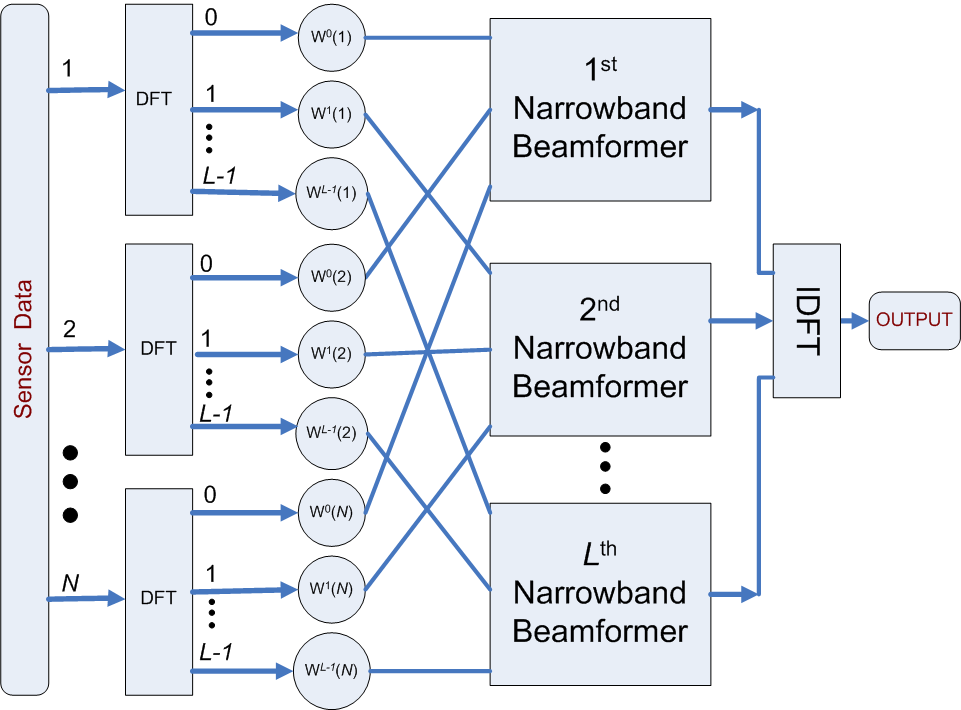}
	\caption{DFT implementation of  wideband  beamforming.}
	\label{labelfig3}
\end{figure}
\begin{equation} \label{j}
\begin{aligned}
 \underset{\mathbf{w} \in \mathbb{C}^{NL}}{\text{minimize}} & \quad   \mathbf{ w}^H\mathbf{R}\mathbf{ w}\\
\text{s.t.} & \quad     \mathbf{ w}^H\mathbf{R}_{s}\mathbf{ w} \geq 1
\end{aligned}
\end{equation}
The equality constraint is relaxed in (\ref{j}) due to the inclusion of the constraint autocorrelation matrix as part of   the objective function, thereby the optimal solution always converges at the equality constraint. The analytical solution of the above  optimization problem exists and  is given by $\mathbf{w}_o = \mathscr{P} \{  \mathbf{R}_{n}^{-1} \mathbf{R}_s  \}=\mathscr{P} \{  \mathbf{R}^{-1} \mathbf{R}_s  \}$. The operator $\mathscr{P} \{. \}$  computes the principal eigenvector of it's argument. Substituting $\mathbf{w}_o= \mathscr{P} \{  \mathbf{R}_{n}^{-1} \mathbf{R}_s  \}  $ into  the SINR formula  yields the corresponding optimum output $SINR_o$ ($\Lambda_{max}$ denotes  the maximum eigenvalue of the matrix);
\begin{equation}  \label{k}
SINR_o=\frac {\mathbf{w}_o^H \mathbf{R}_s \mathbf{w}_o} { \mathbf{w}_o^H \mathbf{R}_{n} \mathbf{w}_o} = \Lambda_{max}\{\mathbf{R}_{n}^{-1} \mathbf{R}_s\},
\end{equation}
which  shows  that the optimum beamformer for maximizing SINR is directly related to the desired and interference plus noise correlation matrices.  
\subsection{DFT Implementation scheme}
Figure \ref{labelfig3} shows   the DFT implementation scheme of   wideband array processing. The received signal $\mathbf{x}(n)$ is processed in the spectral domain by taking an $L$ point DFT for the data received by $k$th sensor ${x_k}(n)$, 
\begin{equation} \label{l}
{X}_k^{(l)} = \sum_{p=0}^{L-1}x_k(n-p)(e^{-j\frac{2\pi}{L}})^{lp}, \, \, \, \, \, \, l \in \{0,1, . . ., \, L-1\}
\end{equation}
Define a vector ${\mathbf{X}}^{(l)} \in \mathbb{C}^N$, containing the  $l$th DFT bin data corresponding to each sensor (superscript $^{(l)}$ denotes the $l$th DFT bin), 
\begin{equation} \label{m}
{\mathbf{X}}^{(l)}= [{X}_1^{(l)}, {X}_2^{(l)}, . . ., {X}_N^{(l)} ]^T
\end{equation}
These  samples are  then combined linearly by the weight vector $\mathbf{w}^{(l)}  \in \mathbb{C}^N$ such that,
\begin{equation}  \label{n}
{y}^{(l)} = {\mathbf{w}^{(l)}}^H \mathbf{X}^{(l)},  \, \, \, \, \, \, l \in \{0,1, . . ., \, L-1\}
\end{equation}
Subsequently, the overall beamformer output $y$ is generated by taking the inverse DFT of $y^{(l)}$ across the $L$ beamformers.
 The DFT implementation scheme seeks to maximum the output SINR for each frequency bin, yielding the optimum beamforming weight vector $\mathbf{w}_{o}^{(l)}$ as the solution of the  following optimization problem,
  \begin{equation}  \label{o}
 \begin{aligned}
 \underset{\mathbf{w}^{(l)}\in \mathbb{C}^N}{\text{minimize}} & \quad   \sum_{l=0}^{L-1}{\mathbf{ w}^{(l)}}^H\mathbf{R}^{(l)}\mathbf{ w}^{(l)}\\
 \text{s.t.} & \quad     {\mathbf{ w}^{(l)}}^H\mathbf{R}_{s}^{(l)}\mathbf{ w}^{(l)} \geq 1.  \quad l \in \{0,1, . . ., \, L-1\}
 \end{aligned}
 \end{equation}
 The correlation matrix $\mathbf{R}^{(l)}=\mathbf{X}^{(l)}{\mathbf{X}^{(l)}}^H$ is the received  correlation matrix for the  $l$th processing bin. Similarly, the source  correlation matrix $\mathbf{R}_{s}^{(l)}$ for the desired source impinging from direction of arrival $\theta_s$ is given by,
 \begin{equation} \label{p}
\mathbf{R}_{s}^{(l)}=\mathbf{S}^{(l)}{\mathbf{S}^{(l)}}^H={\sigma_{s}^{(l)}}^2\mathbf{a}_{\theta_s}^{(l)}{\mathbf{a}_{\theta_s}^{(l)}}^H 
 \end{equation}
 Here, $\mathbf{S}^{(l)}$ is the received data vector representing the desired source and ${\sigma_{s}^{(l)}}^2 $ denotes the power of this source in the $l$th  bin,   $\mathbf{a}_{\theta_s}^{(l)}$  is  the corresponding  steering vector for the source (DOA $\theta_s$) and is defined as follows,
 \begin{equation} \label{q}
 \begin{aligned}
 \mathbf{a}_{\theta_s}^{(l)}= {}&[1 \,  \,  \, e^{j \pi( \frac{\Omega_{min}+l\Delta_{\omega}}{\Omega_{max}}) cos(\theta_s)}  {}\\
  &\,  . \,   . \,  . \, \,  \,  \, \,  \,  \, e^{j \pi( \frac{\Omega_{min}+l\Delta_{\omega}}{\Omega_{max}} ) (N-1)cos(\theta_s)}]^T
 \end{aligned}
 \end{equation}
 Eq. (\ref{q}) models the steering vector for the $l$th DFT bin, where $\Omega_{min}$ is the lower edge of the passband  and $\Delta_{\omega}=\frac{2w_{max}}{L}$ is the spectral resolution. 
The overall output SINR is given by averaging  the SINR over all DFT bins.  Similar to the TDL,  the DFT implementation scheme determines the  optimum sparse array geometry for enhanced MaxSINR performance as explained in the following section.
\section{Optimum sparse array design}  \label{Optimum sparse array design}
The problem of maximizing the principal eigenvalue for the $P$ sensor selection is a combinatorial optimization problem. In this section, we  assume that the   full array data correlation matrix is known. However,  Section \ref{sec4} explains the means to   avoid this assumption through fully augmentable sparse array design or utilizing the matrix completion approach \cite{AMIN20171, HAMZA2020102678, 8682266}. We first formulate the sparse array design for maximizing  SINR in the case of wideband beamforming  as an SDR. Owing to the high computational complexity of  SDR, the problem is solved by SCA, for both the TDL and DFT implementation  schemes.
 
 \subsection{Semidefinite relaxation (SDR) for sparse solution}
 \subsubsection{TDL Implementation scheme}
  We assume that the sensor configuration remains the same within the observation time.  In radar, this assumption amounts to selecting the same $P$ sensors over the coherent processing  interval (CPI).  Therefore, the task is to select $P$ entries from the first $N$ elements of $\mathbf{w}$,  and  the same $P$ entries from each subsequent block of $N$ elements (there are $L$ such blocks). Define $\mathbf{w}_k=[\mathbf{w}(k), \mathbf{w}(k+N), \cdot \cdot \cdot, \mathbf{w}(k+N(L-1))] \in  \mathbb{C}^L$ $(k \in \{1, . . ., \, N\})$ as the weights corresponding to TDL of $k$th sensor.   Then,  in seeking sparse solution, the problem  in  (\ref{j}) can be reformulated as follows,
\begin{equation} \label{a2}
\begin{aligned}
 \underset{\mathbf{w \in \mathbb{C}}^{NL}}{\text{minimize}} & \quad   \mathbf{ w}^H\mathbf{R}\mathbf{w} + \bar{\mu}(\sum_{k=1}^{N}||\mathbf{w}_k||_q)\\
\text{s.t.} & \quad    \mathbf{ w}^H\mathbf{R}_{s}\mathbf{ w} \geq 1 \\
\end{aligned}
\end{equation}
Here,  $||.||_q$ denotes the $q$-norm of the vector, $\bar{\mu}$ is the sparsity regularization parameter. The mixed $l_{1-q}$ norm regularization  is known to thrive  the group sparsity in the solution. In so doing, the above formulation enables the  decision making on whether to collectively select or discard  all the $L$ sampling instances associated with the $k$th  sensor.  Therefore,  structured group sparsity is essential for wideband beamformer, since the final sparse solution has  to ensure that only $PL$  out of $NL$ spatio-temporal possible sampling instances are chosen through only  $P$ physical  sensors.   The relaxed  problem expressed  in the above equation induces  group sparsity in the optimal weight vector  without placing a hard constraint on the desired cardinality. 
 The  constrained minimization in $(\ref{a2})$  can be  penalized instead  by the weighted $l_1$-norm function  to further promote   sparse solutions \cite{6663667, article1, Candes2008},
\begin{equation} \label{b2}
\begin{aligned}
 \underset{\mathbf{w \in \mathbb{C}}^{NL}}{\text{minimize}} & \quad   \mathbf{ w}^H\mathbf{R}\mathbf{w} + \bar{\mu}(\sum_{k=1}^{N}\mathbf{u}^i(k)||\mathbf{w}_k||_q)\\
\text{s.t.} & \quad    \mathbf{ w}^H\mathbf{R}_{s}\mathbf{ w} \geq 1,  \\
\end{aligned}
\end{equation}
where, $\mathbf{u}^i(k)$  is the $k$th element of  re-weighting  vector $\mathbf{u}^i$ at  the $i$th iteration. We choose the  $\infty$-norm  for the  $q$-norm and  replace the weighted $l_1$-norm function   in $(\ref{b2})$  by the  $l_1$-norm  squared  function  with a modified regularization parameter ($\mu$ instead of $\bar{\mu}$). This change does not affect the regularization property of the  $l_1$-norm  \cite{6477161}. The result is,
\begin{equation} \label{c2}
\begin{aligned}
 \underset{\mathbf{w \in \mathbb{C}}^{NL}}{\text{minimize}} & \quad   \mathbf{ w}^H\mathbf{R}\mathbf{w} +   \mu(\sum_{k=1}^{N}\mathbf{u}^i(k)||\mathbf{w}_k||_\infty)^2\\
\text{s.t.} & \quad    \mathbf{ w}^H\mathbf{R}_{s}\mathbf{ w} \geq 1 \\
\end{aligned}
\end{equation}
The SDR of the above problem is realized  by substituting $\mathbf{W}=\mathbf{w}\mathbf{w}^H$. The quadratic function, $ \mathbf{ w}^H\mathbf{R}\mathbf{w}= $ Tr$(\mathbf{ w}^H\mathbf{R}\mathbf{w})= $Tr$(\mathbf{R}\mathbf{w}\mathbf{ w}^H)=$ Tr$(\mathbf{R}\mathbf{W})$, where Tr(.) is the trace of the matrix. Similarly, we replace the regularization term in  (\ref{c2}) by Tr$(\mathbf{U}^i\mathbf{\tilde{W}})$,  with  $\mathbf{U}^i=\mathbf{u}^i(\mathbf{u}^i)^T$  and $\mathbf{\tilde{W}}$ being the auxiliary matrix implementing $\infty-$norm as follows \cite{Bengtsson99optimaldownlink, 5447068, 6477161},
\begin{equation} \label{d2}
\begin{aligned}
\underset{\mathbf{W \in \mathbb{C}}^{NL\times NL}, \mathbf{\tilde{W} \in \mathbb{R}}^{N\times N}}{\text{minimize}} & \quad   \text{Tr}(\mathbf{R}\mathbf{W}) + \mu \text{Tr}(\mathbf{U}^i\mathbf{\tilde{W}})\\
\text{s.t.} & \quad    \text{Tr}(\mathbf{R}_s\mathbf{W}) \geq 1,  \\
& \quad \mathbf{\tilde{W}} \geq  |\mathbf{W}_{ll}|,  \quad l \in \{0, . . ., \, L-1\}   ,\\
& \quad   \mathbf{W} \succeq 0, \, \text{Rank}(\mathbf{W})=1
\end{aligned}
\end{equation}
\begin{algorithm}[t!] \label{algorithm}
	
	\caption{SDR for  the sparse  wideband beamformer}
	
	\begin{algorithmic}[]
		
		\renewcommand{\algorithmicrequire}{\textbf{Input:}}
		
		\renewcommand{\algorithmicensure}{\textbf{Output:}}
		
		\REQUIRE Sparse correlation matrices $\overset{\circ}{\mathbf{R}}$ for TDL-SDR ($\overset{\circ}{\mathbf{R}}{^{(l)}}$ for DFT-SDR), $N$, $P$, $L$, ${\mathbf{R}_s}$ for TDL-SDR (${\mathbf{R}_{s}^{(l)}}$ for DFT-SDR). \\
		
	\ENSURE   MaxSINR beamformer against   $P$ active sensors.   \\
		\textbf{Matrix Completion:} \\
		 Estimate full data receive correlation matrix $\mathbf{\hat{R}}$ for  TDL-SDR  ($\mathbf{\hat{R}}^{(l)}$ for DFT-SDR).\\
		\textbf{Initialization:} \\
		Initialize  $\epsilon$, $\mu_{max}$ (upper limit of binary search), $\mu_{min}$  (lower limit of binary search),  $\mathbf{U}=$ all ones matrix. \\
	
        Appropriate value of  $\mu$ is selected through  the binary search  to ensure $P$ sensor selection.
		\WHILE {(Beamforming weight vector is not $P$ sparse) }
		\STATE Select $\mu=\frac{1}{2}(\mu_{max}+\mu_{min})$ (Binary search).\\
		\STATE   Run the SDR of  (\ref{e2}) for  TDL-SDR  or (\ref{f2}) for DFT-SDR (Use either $\mathbf{\hat{R}}$,$\mathbf{\hat{R}}^{(l)}$   in-lieu of $\mathbf{{R}}$,$\mathbf{{R}}^{(l)}$).\\ 
		\STATE   Update  $\mathbf{U}^i$ according to (\ref{h2}). \\
		Update $\mu_{max}/\mu_{min}$ according to the  binary search. \\
		\ENDWHILE \\
		%
		%
		%
		%
		
		\STATE Run SDR for reduced size correlation matrix corresponding to $P$ sensors of $\mathbf{\tilde{W}}$  and $\mu=0$,  yielding, $\mathbf{w}_o=\mathscr{P} \{  \mathbf{W} \}$ for  TDL-SDR ($\mathbf{w}_{o}^{(l)}=\mathscr{P} \{  \mathbf{W}^{(l)} \}$ for DFT-SDR). 
		
		\RETURN Optimal Beamformer $\mathbf{w}_o$-TDL-SDR ($\mathbf{w}_{o}^{(l)}$-DFT-SDR)
		
		
	\end{algorithmic}
	\label{algorithm1}
\end{algorithm}
The operator  `$|.|$' returns the element wise absolute values of the entries of the matrix, `$\geq$' is the element wise comparison and `$\succeq$' represents the generalized matrix inequality, implementing the positive semidefinite constraint, $\mathbf{W}_{ll} \in \mathbb{C}^{N\times N}$ is the $l$th diagonal block matrix  of $\mathbf{W}$.  We note that the solution matrix $\mathbf{W}$ is Hermitian and therefore, it is sufficient to  constrain the upper or lower triangular  entries of $\mathbf{W}_{ll}$ while forcing $\mathbf{\tilde{W}}$ to be symmetric matrix. In so doing, we reduce the inequality constraints and decrease the run time substantially. In addition, we drop  rank constraint in   (\ref{d2}) which is non convex,    resulting in the following  SDR,
\begin{align} \label{e2}
\underset{\mathbf{W \in \mathbb{C}}^{NL\times NL}, \mathbf{\tilde{W} \in \mathbb{R}}^{N\times N}}{\text{minimize}} & \quad   \text{Tr}(\mathbf{R}\mathbf{W}) + \mu \text{Tr}(\mathbf{U}^i\mathbf{\tilde{W}}), \nonumber \\
\text{s.t.} & \quad    \text{Tr}(\mathbf{R}_s\mathbf{W}) \geq1, \nonumber \\
& \quad \overset{\bigtriangleup}{\mathbf{\tilde{W}}} \geq  | \overset{\bigtriangleup}{\mathbf{W}}_{ll}|,   \quad l \in \{0,1, . . ., \, L-1\}  , \nonumber\\
& \quad \mathbf{\tilde{W}}=\mathbf{\tilde{W}}^T,  \quad \mathbf{W} \succeq 0.
\end{align}
Here,  $^{\bigtriangleup}$  represents the upper or lower triangular entries of the matrix.
 \subsubsection{DFT Implementation scheme}
 The DFT implementation scheme for sparse array design is achieved by starting from   (\ref{o}) and following  similar steps of the TDL. The group sparsity is invoked as the data in each DFT bin requires the underlying array configuration to remain the same for processing over  all DFT bins.  The SDR is finally realized as follows, 
 \begin{align}\label{f2}
 \underset{\mathbf{W}^{(l)} \mathbf{\in \mathbb{C}}^{N\times N}, \mathbf{\tilde{W} \in \mathbb{R}}^{N\times N}}{\text{minimize}} &    \sum_{l=0}^{L-1}\text{Tr}(\mathbf{R}^{(l)}\mathbf{W}^{(l)}) + \mu \text{Tr}(\mathbf{U}^i\mathbf{\tilde{W}}) \nonumber\\
 \text{s.t.} &  \, \,  \text{Tr}(\mathbf{R}_{s}^{(l)}\mathbf{W}^{(l)}) \geq1, \, \,l \in \{0,1, . . ., \, L-1\}, \nonumber\\  
 &  \, \,  \overset{\bigtriangleup}{\mathbf{\tilde{W}}}{^{(l)}} \geq  |\overset{\bigtriangleup}{\mathbf{W}}{^{(l)}}|, \quad l \in \{0,1, . . ., \, L-1\}, \nonumber\\
&   \, \, \mathbf{\tilde{W}}=\mathbf{\tilde{W}}^T,  \quad \mathbf{W} \succeq 0.
 \end{align}
 It is evident from   (\ref{e2}) and (\ref{f2})  that the dimensionality of the TDL implementation scheme is $NL\times NL$, whereas the DFT approach involves $L$ unknown variables of dimensions $N\times N$. 
 \subsubsection{Unit rank promoting iteration}  \label{Unit rank promoting iteration}
 Promoting sparse solutions iteratively would depend on   careful selection of the regularization weighting matrix at each iteration. As suggested in \cite{Candes2008, 6477161}, the weighting matrix  $\mathbf{U}^i$ in  (\ref{e2}) and (\ref{f2})  is initialized unweighted, i.e., a matrix of all ones. Afterwards, this matrix is iteratively updated in an inverse relationship to $\mathbf{\tilde{W}}$, which is related to    the solution matrix $\mathbf{W}$ from the previous iteration. This enables the beamforming weights  having relatively lower magnitude to be   penalized aggressively, hence encouraging them to go to zero in an iterative fashion. The parameter $\epsilon$ prevents  the unwanted case of division by zero and also avoids the solution to converge to  local minima as follows,
\begin{equation} \label{g2}
\mathbf{U}^{i+1}(m,n)=\frac{1}{\mathbf{\tilde{W}}^i(m,n)+\epsilon}
\end{equation}
However,   the solution matrix $\mathbf{W}$ in the case of the TDL implementation scheme is not exactly rank  one  at initial iterations. The problem aggravates when the  weight matrix is  updated according to    the above equation, inadvertently pushing the  rank of $\mathbf{W}$ to build up with each iteration. In this respect, updating $\mathbf{U}$ according to (\ref{g2}) favors solutions of higher ranks, and, as such, fails to yield  sparse configurations iteratively. To mitigate this problem, we  propose a modified re-weighting scheme that relies on updating the regularization weighting matrix  that depends on the inverse of a  rank $1$ matrix $\mathbf{Y}$ instead of   $\mathbf{\tilde{W}}$ as follows, 
\begin{equation} \label{h2}
\mathbf{U}^{i+1}(m,n)=\frac{1}{\mathbf{Y}^i(m,n)+\epsilon},
\end{equation}
where, $\mathbf{Y}^i=\mathbf{y}^i(\mathbf{y}^i)^T,$ for $\mathbf{y}^i=\frac{1}{L}\sum_{l=1}^{L}(\mathbf{|\mathscr{P}\{W}_{ll}^i\}|)^2 $.  Clearly, $\mathbf{Y}^i$ is a rank one matrix that is synthesized from the rank one approximation of each block diagonal matrix $\mathbf{{W}}_{ll}^i$. It is noted that  $\mathbf{{W}}_{ll}^i$ are the only entries of $\mathbf{{W}}$ that are constrained  by the SDR formulated in (\ref{e2}). Since  $\mathbf{{W}}_{ll}^i$ are the diagonal block matrices of  the solution matrix $\mathbf{{W}}$, then sparsity is implicitly  encouraged in $\mathbf{{W}}$ by unit rank penalization.    This modified re-weighting approach  given by  (\ref{h2}) effectively solves the  optimum sparse array  selection problem for the wideband beamforming.  It is noted that the arbitrarily chosen sparsity parameter $\mu$ does not ensure  the   final solution to be exactly $P$ sparse. In order to do so, the optimization problem should be solved  against various values of $\mu$. This is typically achieved by successively running the optimization and updating the  values of $\mu$  through a binary search  over the possible upper and lower limit of $\mu_{max}/\mu_{min}$, until the solution converges to $P$ sensors \cite{6477161}.   The proposed algorithm for achieving  the sparse  optimal weight vector  under the TDL and DFT implementation schemes is summarized in  Algorithm 1. 
\begin{algorithm}[t!] \label{algorithm}
	
	\caption{SCA   for the sparse wideband beamformer}
	
	\begin{algorithmic}[]
		
		\renewcommand{\algorithmicrequire}{\textbf{Input:}}
		
		\renewcommand{\algorithmicensure}{\textbf{Output:}}
		
		\REQUIRE Sparse correlation matrices $\overset{\circ}{\mathbf{R}}$  for  TDL-SCA ($\overset{\circ}{\mathbf{R}}{^{(l)}}$ for DFT-SCA), $N$, $P$, $L$, ${\mathbf{R}_s}$ for TDL-SCA (${\mathbf{R}_{s}^{(l)}}$ for DFT-SCA). \\
	
	\ENSURE   MaxSINR beamformer against   $P$ active sensors.   \\
	\textbf{Matrix Completion:} \\
	Estimate full data receive correlation matrix $\mathbf{\hat{R}}$ for  TDL-SCA  ($\mathbf{\hat{R}}^{(l)}$ for DFT-SCA).\\
	\textbf{Initialization:} \\
		Initialize  $\mathbf{m}$,  $b$,  $\epsilon$, $\mu_{max}$ (upper limit of binary search), $\mu_{min}$  (lower limit of binary search), $\mu=0$,  $\mathbf{u}^{i}$= all ones vector. 
		\WHILE {(Solution does not converge for $\mu=0$)}
		
		\STATE   Run   (\ref{n2}) for  TDL-SCA or (\ref{o2}) for DFT-SCA.\\ 
		\ENDWHILE \\
		\WHILE {(Beamforming weight vector is not $P$ sparse)}
		\STATE Select $\mu=\frac{1}{2}(\mu_{max}+\mu_{min})$ (Binary search).\\
		\STATE   Run   (\ref{n2}) for  TDL-SCA or (\ref{o2}) for DFT-SCA (Use  $\mathbf{\hat{R}}$ or $\mathbf{\hat{R}}^{(l)}$ in-lieu of $\mathbf{{R}}$, $\mathbf{{R}}^{(l)}$ to synthesize $\mathbf{\tilde{R}}$, $\mathbf{\tilde{R}}^{(l)}$).\\  
		(Also use the optimal non sparse solution from  the previous while loop for  $\mathbf{m}$ and $b$).\\ 
		\STATE   Update the regularization weighting parameter $\mathbf{u}^{i+1}(k)=\frac{1}{||\mathbf{\tilde{w}}_k^i||_2+\epsilon}$.\\
		Update $\mu_{max}/\mu_{min}$ according to the  binary search. \\
		\ENDWHILE \\
		%
		%
		%
		%
		
		\STATE Run  (\ref{n2}) for  TDL-SCA or (\ref{o2}) for DFT-SCA, with reduced dimension  corresponding to $P$ sensors of $\mathbf{\tilde{w}}$  and $\mu=0$,  yielding, optimal weight vector. 
		
		\RETURN Optimal Beamformer $\mathbf{w}_o$-TDL-SCA ($\mathbf{w}_{o}^{(l)}$-DFT-SCA)
	\end{algorithmic}
	\label{algorithm2}
\end{algorithm}\vspace{-3mm}
\subsection{Successive convex approximation (SCA)}
\subsubsection{TDL Implementation scheme}
The problem in (\ref{j}) can equivalently be rewritten by swapping the objective and constraint functions as follows,
\begin{equation} \label{i2}
\begin{aligned}
 \underset{\mathbf{w \in \mathbb{C}}^{NL}}{\text{maximize}} & \quad   \mathbf{ w}^H\mathbf{R}_s\mathbf{ w}\\
\text{s.t.} & \quad     \mathbf{w}^H\mathbf{R}\mathbf{ w} \leq 1
\end{aligned}
\end{equation}
Although this swapping operation allows the associated constraint to be convex, however it renders the objective function non convex. We note that the formulation in (\ref{i2}) doesn't have a trivial solution $\mathbf{w} =0$, as the objective and constraint are coupled due to $\mathbf{R}_s$  being part of $\mathbf{R}$.   The maximization of the convex function  is replaced by the minimization of the concave function. The transformation to the minimization problem will later enable carrying out the minimization based on  $P$ sparse solution. Rewriting (\ref{i2}) by reversing the sign of the desired source correlation matrix  $\bar{\mathbf{R}}_s=-\mathbf{R}_s$ as follows, 
\begin{equation} \label{j2}
\begin{aligned}
 \underset{\mathbf{w \in \mathbb{C}}^{NL}}{\text{minimize}} & \quad   \mathbf{ w}^H\bar{\mathbf{R}}_s\mathbf{ w}\\
\text{s.t.} & \quad     \mathbf{ w}^H\mathbf{R}\mathbf{ w} \leq 1
\end{aligned}
\end{equation}
The beamforming weight vectors are generally complexed valued, whereas the  quadratic functions are real. This observation  allows expressing the problem with  only real variables and  is typically accomplished by  replacing the correlation matrix $\bar{\mathbf{R}}_s$ by $\tilde{\mathbf{R}}_s$  and concatenating the beamforming weight vector accordingly \cite{Ibrahim2018MirrorProxSA}, \vspace{-2mm}
\begin{multline} \label{k2}
\tilde{\mathbf{R}}_s=\begin{bmatrix}
\text{real}(\bar{\mathbf{R}}_s)       & -\text{imag}(\bar{\mathbf{R}}_s)   \\
\\
\text{imag}(\bar{\mathbf{R}}_s)       & \text{real}(\bar{\mathbf{R}}_s)   \\
\end{bmatrix},
\tilde{\mathbf{w}}=\begin{bmatrix}
\text{real}({\mathbf{w}})       \\
\\
\text{imag}({\mathbf{w}})   \\
\end{bmatrix}
\end{multline}
Similarly, the received data correlation matrix $\mathbf{R}$ is replaced by $\tilde{\mathbf{R}}$. The problem in  (\ref{j2}) then  becomes,
\begin{equation} \label{l2}
\begin{aligned}
 \underset{\mathbf{\tilde{w} \in \mathbb{R}}^{2NL}}{\text{minimize}} & \quad   \mathbf{\tilde{ w}}^{T}\tilde{\mathbf{R}}_s\mathbf{\tilde{ w}}\\
\text{s.t.} & \quad     \mathbf{\tilde{ w}}^{T}\mathbf{\tilde{R}}\mathbf{\tilde{w}} \leq 1
\end{aligned}
\end{equation}
We can then proceed to convexify the objective function  by utilizing  the first order approximation  iteratively,
 \begin{equation} \label{m2}
 \begin{aligned}
 \underset{\mathbf{\tilde{ w} \in \mathbb{R}}^{2NL}}{\text{minimize}} & \quad   {\mathbf{m}^i}^{T}\mathbf{\tilde{ w}}+b^i\\
 \text{s.t.} & \quad     \mathbf{\tilde{ w}}^{T}\mathbf{\tilde{R}}\mathbf{\tilde{w}} \leq 1,
 \end{aligned}
 \end{equation}
The linearization coefficients  $\mathbf{m}^i$ and ${b^i}$  are updated as, $\mathbf{m}^{i+1}=2\tilde{\mathbf{R}}_s\mathbf{\tilde{ w}}^i$, and  $b^{i+1}=-{\mathbf{\tilde{w}}^i}^{T}\tilde{\mathbf{R}}_s \mathbf{\tilde{w}}^i $ (superscript $^i$ denotes the iteration number). Finally, to invoke sparsity in the beamforming weight vector, the re-weighted mixed $l_{1-2}$ norm is adopted primarily for promoting group sparsity,  \vspace{-3mm}
 \begin{equation} \label{n2}
 \begin{aligned}
  \underset{\mathbf{\tilde{w} \in \mathbb{R}}^{2NL}}{\text{minimize}} & \quad   {\mathbf{m^i}^{T}}\mathbf{\tilde{ w}}+b^i + \mu(\sum_{k=1}^{N}\mathbf{u}^i(k)||\mathbf{\tilde{w}}_k||_2)\\
 \text{s.t.} & \quad     \mathbf{\tilde{ w}}^{T}\mathbf{\tilde{R}}\mathbf{\tilde{ w}}\leq 1
 \end{aligned}
 \end{equation}
 Here, $\mathbf{\tilde{w}}_k \in \mathbb{R}^{2L}$ are the beamforming weights corresponding to TDL of $k$th sensor. Discouraging a sensor ($||.||_2$ denotes the $l_2$ norm) implies a simultaneous removal of both the real and corresponding imaginary entries of all beamforming weights  associated with the removed sensor \cite{Ibrahim2018MirrorProxSA}. 
\subsubsection{DFT implementation scheme}
The above formulation can be  extended  for the DFT  implementation scheme as follows:\vspace{-3mm}
\begin{equation} \label{o2}
\begin{aligned}
 \underset{\mathbf{\tilde{w}}^{(l)} \mathbf{\in \mathbb{R}}^{2N}}{\text{minimize}} &    \sum_{l=0}^{L-1}( {{\{\mathbf{m}^{(l)}}\}^i}^{T}\mathbf{\tilde{ w}}^{(l)}+\{b^{(l)}\}^i) + \mu(\sum_{k=1}^{N}\mathbf{u}^i(k)||\mathbf{\tilde{w}}_k||_2)\\
\text{s.t.} & \quad     {\mathbf{\tilde{ w}}^{(l)}}^{T}\mathbf{\tilde{R}}^{(l)}\mathbf{\tilde{ w}}^{(l)} \leq 1, \quad   l \in \,\, \{0, 1, . . .  L-1\}, 
\end{aligned}
\end{equation}
where $\mathbf{\tilde{w}}_k \in R^{2L}$ contains  the $L$ DFT bins data for the $k$th sensor, $\{{\mathbf{m}^{(l)}}\}{^i}$ and $\{{b^{(l)}\}{^i}}$ are the approximation coefficients at the $i$th iteration for the desired source correlation matrix in the $l$th bin, with $\{{\mathbf{m}^{(l)}}\}{^i}=2\tilde{\mathbf{R}}_{s}^{(l)}\{{\mathbf{\tilde{ w}}^{(l)}}\}{^{i}},  \{{b^{(l)}\}{^i}}=-{\{\mathbf{\tilde{w}}^{(l)}\}^i}^{T}\tilde{\mathbf{R}}_{s}^{(l)}\{\mathbf{\tilde{w}}^{(l)}\}^i$.
The initial estimates $\{{\mathbf{m}^{(l)}}\}{^i}$ and $\{{b^{(l)}\}{^i}}$  are calculated for the optimal non sparse solution. These parameters can be found by setting the sparsity parameter  $\mu$ to zero. In  so doing,  the solution and the  corresponding parameters $\{{\mathbf{m}^{(l)}}\}{^i}$ and $\{{b^{(l)}\}{^i}}$    converge to the optimal value against  the full array elements. Using these values as initial conditions  has proven  appropriate in our design for  recovering effective sparse solutions.  The sparsity parameter $\mu$ is  chosen according to the binary search over the possible range of $\mu$ to warrant the desired cardinality of the beamforming weight vector as explained before in section \ref{Unit rank promoting iteration}. The $k$th entry of re-weighting vector $\mathbf{u}^i(k)$ is updated according to $\mathbf{u}^{i+1}(k)=\frac{1}{||\mathbf{\tilde{w}}_k^i||_2+\epsilon}$. The SCA for sparse array design for wideband beamforming is summarized in  Algorithm 2.

\subsection{Computational complexity}
In general, QCQP of order $M$ with $T$ quadratic constraints can   be solved to an arbitrary small accuracy  $\zeta$ by employing  interior point methods  involving   the   worst case polynomial complexity of  $\mathcal{O}\{$max$(T,M)^4M^{(1/2)}\log(1/\zeta)\}$   \cite{5447068}. It is apparent from   (\ref{e2}) and (\ref{f2})  that the order of the TDL implementation scheme is $NL$, whereas  the DFT approach involves $L$ unknown variables, each of order $N$. Therefore, the polynomial complexity for TDL implementation scheme is  $\mathcal{O}\{(NL)^{4.5}\log(1/\zeta)\}$, and is  $\mathcal{O}\{N^{4.5}\log(1/\zeta)\}$ (assuming $N>L$) for the DFT implementation scheme. This renders the latter computationally viable. The polynomial complexity  of  SCA for TDL implementation scheme is  $\mathcal{O}\{(NL)^{3}\log(1/\zeta)\}$, and is  $\mathcal{O}\{N^{3}\log(1/\zeta)\}$  for the DFT implementation.  Hence, polynomial complexity is considerably lower for SCA as compared to the SDR formulation, as the latter  intrinsically  squares the number of   variables involved, essentially exacerbating the  runtime \cite{Ibrahim2018MirrorProxSA}. 

\section{Sparse  matrix completion of block Toeplitz matrices} \label{sec4}
The aforementioned  sparse array design formulations   require    the received data correlation matrix corresponding to the full array aperture. This  is a  rather stringent requirement  in an adaptive switching environment  where the data is fetched  from  only $P$ active  sensor locations over a given observation period.  The received data correlation matrix for the sparse array design using TDL implementation scheme has $L^2$($N^2-P^2$) missing correlation entries, whereas there  are $L$($N^2-P^2$) missing correlation values for the DFT implementation scheme. Clearly, for the large values of $L$, the TDL implementation scheme has significantly higher number of missing correlation entries as compared to the DFT implementation scheme. 

Recently, the hybrid sparse design for the  narrowband beamforming was introduced to alleviate the issue of  missing correlation lags in the received data correlation matrix \cite{8892512}.  This is primarily  achieved by pre-allocating few sensors  to guarantee a fully augmentable sparse array, while engaging the remaining degrees of freedom (DOF) to maximize the SINR.  However, locking in few DOFs to ensure the array full augmentability can lead to suboptimal sparse beamformers. Alternatively, the matrix completion    approach can be used to provide the missing lags \cite{Abramovich:1999:PTC:2197953.2201686, 709534, 7935534}.  We propose, herein,  sparse matrix completion to efficiently exploit the structure of the data correlation matrix to recover the missing correlation values. The received data correlation matrix for the TDL implementation scheme is a Hermitian positive definite matrix but it also follows a block Toeplitz formation, as shown below,
\begin{multline} \label{a4}
\mathbf{R}=\begin{bmatrix}
\mathbb{T}_0       & {\mathbb{T}}_1  & {\mathbb{T}}_2 & \ldots & {\mathbb{T}}_{L-1} \\
\\
\mathbb{T}_{-1}       & {\mathbb{T}}_0    & {\mathbb{T}}_1  & \ldots   & {\mathbb{T}}_{L-2}  \\
\\
{\mathbb{T}}_{-2}       & {\mathbb{T}}_{-1}    & {\mathbb{T}}_0  & \ldots  & {\mathbb{T}}_{L-3}  \\
\\
\vdots       & \vdots    & \vdots  & \ddots  & \vdots   \\
\\
{\mathbb{T}}_{-(L-1)}       & {\mathbb{T}}_{-(L-2)}    & {\mathbb{T}}_{-(L-3)}  & \ldots  & {\mathbb{T}}_0  \\
\end{bmatrix}
\end{multline}
By definition, block Toeplitz matrices  doesn't necessitate each comprising block  to be Toeplitz within itself. Therefore, matrix $\mathbf{R}$ in (\ref{a4}) represents a special case of block Toeplitz matrices, where the Toeplitz structure  is also preserved for each constituent block $\mathbb{T}_k \in \mathbb{C}^{N\times N}$. Because of  the matrix Hermitian symmetry, we also have $\mathbb{T}_k^{H}=\mathbb{T}_{-k} (\text{for} \, \, \, k\neq 0 $). 
Instead of recovering $\mathbf{R}$ as a single unit, we focus on  completing  the constituent blocks and then synthesizing the full correlation matrix $\mathbf{R}$. This approach  potentially caps the computational expenses considerably but also efficiently exploits the formation of $\mathbf{R}$.

There is an important distinction between the constituent blocks $\mathbb{T}_0$ and $\mathbb{T}_k \, \, \,  (\text{for} \, \, \, k\neq 0 $). It is noted that $\mathbb{T}_0$ is positive definite Hermitian Toeplitz matrix, whereas $\mathbb{T}_k$ $ \, \, \,  (\text{for} \, \, \, k\neq 0 $) are indefinite Toeplitz matrices which  are not necessarily Hermitian. Therefore, we  resort to  two different ways with regards to our treatment of $\mathbb{T}_0$ and $\mathbb{T}_k$ while adopting a  Toepltiz  matrix completion scheme  under  the low rank assumption. It is  known that the correlation matrix $\mathbf{R}$  for the  wideband far field  sources impinging on the ULA resultantly follows the structure in (\ref{a4}) and can be represented effectively with  a relatively  low rank approximation depending on the observed source  time-bandwidth product \cite{1143832}.   The  trace heuristic is a well known approach which is  adopted generally as a convex surrogate in recovering low rank matrices. This approach has been successfully used in many areas of control systems and array processing to recover simpler and low rank data models \cite{Recht:2010:GMS:1958515.1958520, Mohan:2012:IRA:2503308.2503351, 5447068}. Moreover, it has been shown  that the trace heuristic is equivalent to the nuclear norm minimization in recovering positive semidefinite correlation matrices\cite{8474369,  7539135, 7935534, 7833203}. The low rank positive semidefinite Toeplitz matrix completion problem has been proposed in \cite{8472789} for interpolating missing correlation lags in coprime array configuration and can be adopted to interpolate $\mathbb{T}_0$ as follows,
\begin{equation} \label{b4}
\begin{aligned}
\underset{l \in \mathbb{C}^{N}}{\text{minimize}} & \quad  ||\mathcal{T}(l)\odot \mathbf{Z}-\overset{\circ}{\mathbb{T}}_0  ||_F^2 + \zeta \text{Tr}(\mathcal{T}(l))\\
\text{s.t.} & \quad   \mathcal{T}(l) \succeq 0
\end{aligned}
\end{equation}
Here,  the unknown Hermitian Toeplitz matrix $\mathcal{T}(l)$, can uniquely be defined by a single  vector $l$ representing  the  first row of  $\mathcal{T}(l)$, and $l^{H}$ denoting the matrix first  column.  Matrix $\overset{\circ}{\mathbb{T}}_0 $ is the received data correlation matrix with the missing correlation values set equal  to zero. The   element wise product is denoted by symbol `$\odot$' and  `$\succeq$' implements the positive  semidefinite constraint. The objective function attempts to minimize  the  error  between the observed correlation values and the corresponding entries of $\mathcal{T}(l)$ implemented through the Frobenius norm of the error matrix (The function `$||.||_F^2$',  represents the square of the Frobenius norm of matrix which returns the sum of square of it's entries).   The parameter `$\zeta$' pursues  the trade off   between the denoising  term and the trace heuristic to recover a simpler low rank model. The nominal value of the  parameter  `$\zeta$'   is challenging to locate and is typically gleaned from the numerical experience. In order to do away with the nuisance parameter  `$\zeta$',   we adopt a fusion based approach more suited to our application. We note that the non zero elements in $\overset{\circ}{\mathbb{T}}_0 $ can be segregated into two classes. With regards to the sparse entries in $\overset{\circ}{\mathbb{T}}_0 $,  either we have the  whole sub-diagonal entries  missing in $\overset{\circ}{\mathbb{T}}_0 $ or the sub-diagonals are sparse. The former situation arises if there is missing correlation lag in $\overset{\circ}{\mathbb{T}}_0 $, whereas the latter arises when the corresponding correlation lag is present but lacking the intrinsic redundancy corresponding to the compact sensor grid. The observed correlation lags are averaged across the sub diagonal to filter the sensor noise as follows, 
 \begin{equation} \label{c4}
  \doublehat{l}^i(k)=(\frac{1}{c_k}) \sum_{\forall \, m-n=k}\overset{\circ}{\mathbb{T}}^i_0 (m,n)    
 \end{equation}
 Here,   $\overset{\circ}{\mathbb{T}}^i_0$  is the $\mathbf{R}(i,i)$ entry in (\ref{a4}) denoting the estimate of $\overset{\circ}{\mathbb{T}}^i_0$ at $i$th sampling instance, $k \in 0,1, \cdots, N-1$ represents the respective lag or the sub diagonal of $\overset{\circ}{\mathbb{T}}^i_0$ and  $c_k$ is the observed redundancy of the $k$th lag. As evident in (\ref{a4}), there are $L$ copies of  $\overset{\circ}{\mathbb{T}}_0$ corresponding to $L$ sampling instances. Hence, $\doublehat{l}^i(k)$ is averaged over $L$ blocks to yield an estimate of the given lag  $\hat{l}(k)=\frac{1}{L}\sum_{i=0}^{L-1}\doublehat{l}^i(k)$. The fused matrix completion formulation, therefore, substitutes the sparse sub-diagonals with the estimated average value $\hat{l}(k)$, whereas the completely missing sub-diagonals are interpolated   as follows,
\begin{equation} \label{d4}
\begin{aligned}
\underset{l \in \mathbb{C}^{N}}{\text{minimize}} & \quad \text{Tr}(\mathcal{T}(l))\\
\text{s.t.} & \quad   l(\text{lag present}) = \hat{l}(k),\\
 & \quad   \mathcal{T}(l) \succeq 0  
\end{aligned}
\end{equation}
The above formulation relies on fairly accurate estimate of the observed correlation lag which  not only involves averaging over the corresponding sub diagonal but also  across $L$ sampling instances. Such accuracy can  become   challenging to meet the semidefinite constraint if the available degrees of freedom are few. To circumvent this problem, the $0$ lag is removed from the constraint which gives the additional degree of freedom to the  algorithm  to set aside an  appropriate loading factor to make the problem feasible. It is noted that the formulation in (\ref{d4}) would choose the  minimum possible  diagonal loading as it requires to minimize the trace heuristic and hence strives to maximize the sparse  eigenvalues.

The trace heuristic can also be extended for indefinite matrices to recover sparse models \cite{Candes2009}. We  couple this observation along with the above discussion  to perform low rank matrix completion for $\mathbb{T}_k \, \, \,  (\text{for} \, \, \, k\neq 0 $) as follows, 
\begin{equation} \label{e4}
\begin{aligned}
\underset{l_r, l_c \in \mathbb{C}^{N}}{\text{minimize}} & \quad \text{Tr}(\mathbf{W}_1) +  \text{Tr}(\mathbf{W}_2)\\
\text{s.t.} & \quad   l_r(\text{lag present}) = \hat{l_r}(k),\\
\text{s.t.} & \quad   l_c(\text{lag present}) = \hat{l_c}(k),\\
     & \quad    
\begin{bmatrix}%
\mathbf{W}_1 & \mathcal{T}(l_r,l_c) \\
\mathcal{T}(l_r,l_c) & \mathbf{W}_2
\end{bmatrix}\succeq 0
\end{aligned} 
\end{equation}
Here, $\mathbf{W}_1$ and $\mathbf{W}_2$ are auxiliary matrices implementing trace heuristic to recover low rank indefinite matrices.  The function  $Toeplitz(l_r,l_c)$ returns the Toeplitz matrix with $l_r$ and $l_c$ being the first row and the first column, respectively. This distinction is important as in general $\mathbb{T}_k \, \, \,  (\text{for} \, \, \, k\neq 0 $) is not a Hermitian Toeplitz matrix. The formulation in (\ref{e4}) is repeated to yield an estimate for all constituent Toeplitz blocks. 

 Upon performing matrix completion for each constituent Toeplitz block, the individual Toeplitz blocks can be plugged back into  (\ref{a4}) to yield an estimate  $\mathbf{\doublehat{{R}}}$. We can improve the estimate  $\mathbf{\doublehat{{R}}}$ by incorporating the   noise variance which  is generally known or estimated apriori. This is  achieved through the maximum likelihood estimate (MLE) approach where the eigenvalues corresponding to the  eigenvectors of the  noise subspace are set equal to the noise floor while the remaining eigenvalues are kept the same. The MLE of $\mathbf{\doublehat{{R}}}$ is donated by $\mathbf{\hat{R}}$ and is given by the outer product of the original eigenvectors of $\mathbf{\doublehat{{R}}}$ reweighted by the modified eigenvalues.  However, it is noted that in practice  the number of  eigenvectors associated with the noise floor are not exactly known. Nevertheless, we only reset those  eigenvalues which are less than the noise floor. Finally, the maximum likelihood estimate  $\mathbf{\hat{R}}$, is used in lieu of $\mathbf{\doublehat{{R}}}$, to carry out the data dependent optimization for MaxSINR. It is also noted that unlike $\mathbf{\doublehat{{R}}}$, the  matrix $\mathbf{\hat{R}}$ is no longer block Toeplitz   but is now guaranteed to be positive definite as is strictly required to implement the proposed sparse optimization algorithms. The formulation in (\ref{d4}) is sufficient for the DFT implementation scheme  which only involves $L$ received data matrices corresponding to the $L$ DFT bins.  
 \section{Simulations} \label{Simulations}
The effectiveness of  the sparse array design for  MaxSINR  beamforming is demonstrated by  design examples considering a wideband source operating in the presence of a mix of narrowband and wideband jammers. The MATLAB-based CVX modeling tool  is used for convex optimization. The  importance  of   sparse array design for MaxSINR is further emphasized  by comparing the  optimum  design with the sub optimum  array configurations under the TDL  and DFT implementation schemes. The simulation results are presented for perspective linear sensor locations, nevertheless, the proposed algorithms are applicable to  rectangular grid points or arbitrary placed sensors on  3D surfaces.
\subsection{Example 1}
The task is to select $P=8$ sensors from $N=20$  possible equally spaced locations with  inter-element spacing of $\lambda_{min}/2$.  We consider   $8$ delay line filter taps associated  with each selected  sensor ($8$ DFT bins for DFT implementation scheme) implying $L=8$. 
A desired wideband   point source   impinges on a linear array from  DOA $40^0$. The fractional bandwidth of the system w.r.t. the center frequency is  $0.22$. The  desired source has uniform PSD (power spectral density),  occupying the normalized frequency spectrum from -0.25 to 0.25 cycles/sample. Three strong wideband jammers  occupying the entire  frequency band from -0.5 to 0.5 cycles/sample, are located at angles  $45^0$, $50^0$ and $140^0$. There is another wideband jammer   at $30^0$, covering the same spectrum as that of the desired source (-0.25 to 0.25 cycles/sample). In addition, a narrowband jammer lock onto the source carrier frequency is  located  at $150^0$.    The SNR of the desired signal is $0$ dB, and the INR of  each interfering signals is set to $30$ dB. 
\begin{figure}[!t]
	\centering
	\includegraphics[width=3.78 in, height=2.10 in]{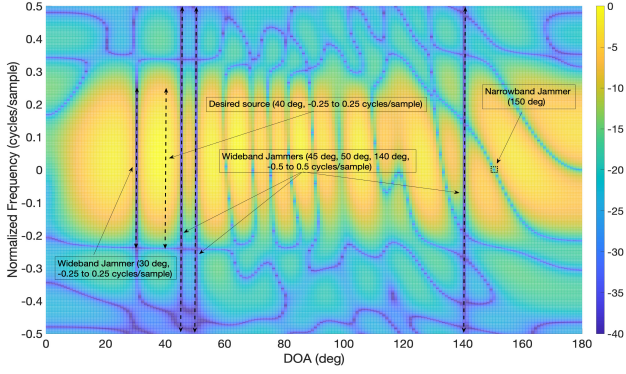}
	\caption{Frequency dependent beampattern for the  array configuration recovered through convex relaxation. }
	\label{labelfig4}
\end{figure}

Figure \ref{labelfig4} shows the frequency dependent beampattern  for the   array configuration recovered through TDL-SDR. It is evident from the beampattern that a  high gain is maintained throughout the band of interest for the desired source signal while alleviating  the interferers from respective  DOAs and  frequency bands. These beampattern characteristics   translate to an output SINR of  $8.96$ dB. The TDL-SDR performs  close to the optimum array found by enumeration which has an output SINR of $9.3$ dB. The corresponding sparse array configurations obtained by enumeration and TDL-SDR are shown in the Fig. \ref{op_enumer_608} and \ref{op_algo_2027}, respectively (the green color denotes the sensor locations selected, whereas the gray color shows the sensor locations not selected). It is important to mention  that  the optimum array found by enumeration requires a search over $125970$ possible sparse array configurations, which has a prohibitive  computational cost attributed to expensive singular value decomposition (SVD) for each enumeration.  The optimum sparse array design achieved through TDL-SCA is shown in  Fig. \ref{op_notap_775}. This array configuration is capable of delivering output SINR of $6.81$ dB with the use of optimal beamforming weights. 
This performance is inferior to that of the sparse array found through TDL-SDR by around $2$ dB, however with   less computational complexity.

\begin{figure}[!t]
	\centering
	\includegraphics[width=3.78 in, height=2.10 in]{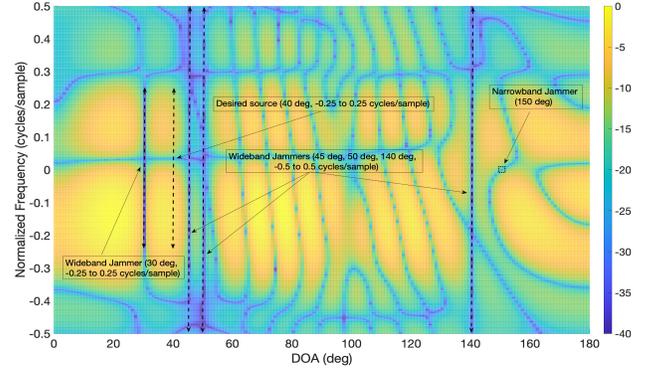}
	\caption{Frequecy dependent beampattern for the array configuration shown in  Fig. \ref{wo_enu_ula}.}
	\label{labelfig5}
\end{figure}
In general, for any given  array configuration,  the TDL implementation scheme yields marginally higher SINR as compared to the DFT implementation scheme. Owing to the reduced computational cost of the DFT sparse design, and relatively better performance of the TDL sparse design, one can entwine TDL and DFT beamformers to capitalize on  the merits of both approaches. That is, we proceed to find the optimum TDL beamformer weights based on the sensor array configuration that results from the optimum DFT-based implementation scheme.  This  refer  a dual-domain design implementation scheme  since it considers both time and frequency domains in generating the optimum receiver design. This design has slightly elevated  computational expense over the  DFT design, as it  involves calculating the  optimum TDL beamformer weights corresponding to the DFT optimized configuration.  For the underlying problem, the dual-domain design, through  the DFT-SDR and DFT-SCA gives an output SINR of $8$~dB and $8.8$~dB respectively, which is close to the   maximum possible SINR of $9.3$~dB.  The sparse array configurations rendered by  DFT-SDR and DFT-SCA  are shown in  Figs. \ref{wo_enu_924} and \ref{wo_enu_925}. Therefore, we remark that for the  example considered, the dual domain design underlying the DFT implementation scheme, performs closely   to the design carried out exclusively by the TDL implementation scheme.  
\begin{figure}[!t]
	\centering
	\begin{subfigure}[b]{0.48\textwidth}
		\includegraphics[height=0.4 in, width=1\textwidth]{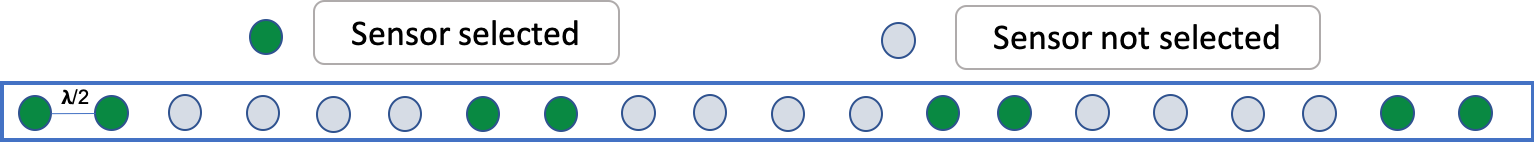}
		\caption{}
		\label{op_enumer_608}
	\end{subfigure}
	\begin{subfigure}[b]{0.48\textwidth}
		\includegraphics[height=0.20in, width=1\textwidth]{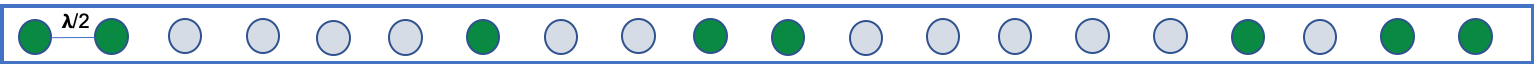}
		\caption{}
		\label{op_algo_2027}
	\end{subfigure}
	\begin{subfigure}[b]{0.48\textwidth}
		\includegraphics[height=0.20in, width=1\textwidth]{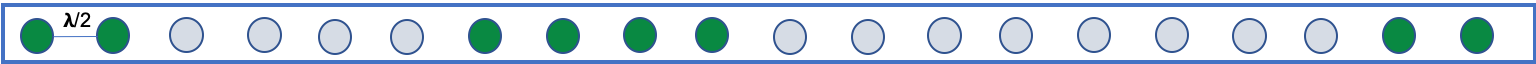}
		\caption{}
		\label{op_notap_775}
	\end{subfigure}
	\begin{subfigure}[b]{0.48\textwidth}
		\includegraphics[height=0.20in, width=1\textwidth]{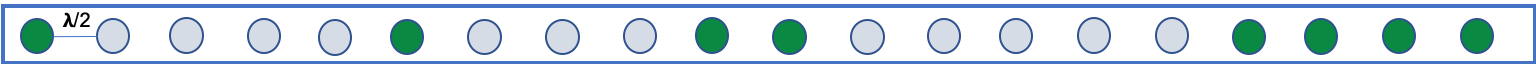}
		\caption{}
		\label{wo_enu_924}
	\end{subfigure}
		\begin{subfigure}[b]{0.48\textwidth}
		\includegraphics[height=0.20in, width=1\textwidth]{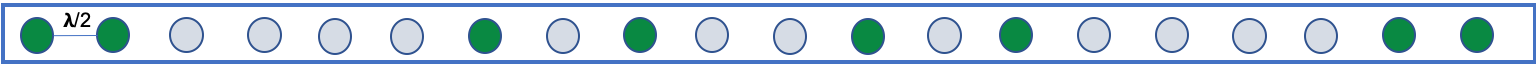}
		\caption{}
		\label{wo_enu_925}
	\end{subfigure} 
	\begin{subfigure}[b]{0.48\textwidth}
		\includegraphics[height=0.20in, width=1\textwidth]{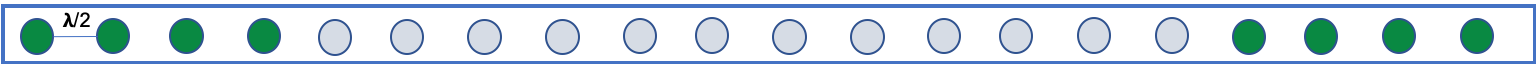}
		\caption{}
		\label{wo_enu_ula}
	\end{subfigure} 
	\caption{Example 1 - (a) Optimum  array TDL implementation scheme (Enumeration)  (b)  TDL-SDR  (c) TDL-SCA  (d) DFT-SDR (e)   DFT-SCA (f)  $8$ sensor centrally sparse array}
	\label{op_c_co_c}
\end{figure}
\begin{figure}[!t]
	\centering
	\begin{subfigure}[b]{0.48\textwidth}
		\includegraphics[height=0.35in, width=1\textwidth]{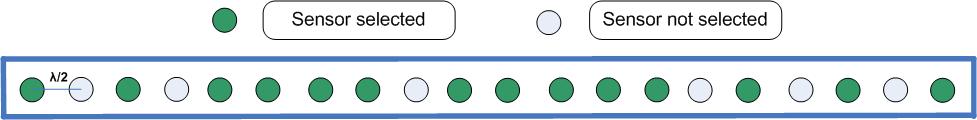}
		\caption{}
		\label{op_enumer_6082}
	\end{subfigure}
	\begin{subfigure}[b]{0.48\textwidth}
		\includegraphics[height=0.20in, width=1\textwidth]{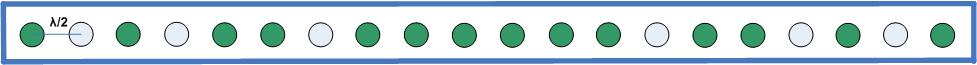}
		\caption{}
		\label{op_notap_7752}
	\end{subfigure}
	\begin{subfigure}[b]{0.48\textwidth}
		\includegraphics[height=0.20in, width=1\textwidth]{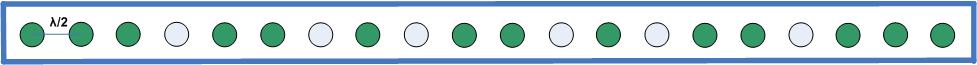}
		\caption{}
		\label{op_algo_20272}
	\end{subfigure}
	\begin{subfigure}[b]{0.48\textwidth}
		\includegraphics[height=0.20in, width=1\textwidth]{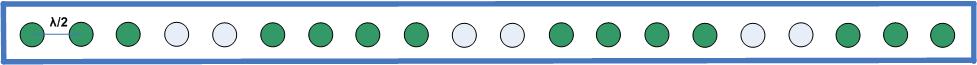}
		\caption{}
		\label{wo_enu_9262}
	\end{subfigure}
	\begin{subfigure}[b]{0.48\textwidth}
		\includegraphics[height=0.20in, width=1\textwidth]{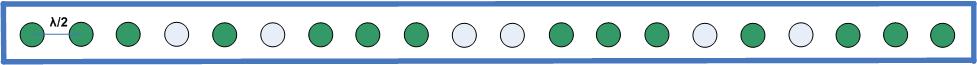}
		\caption{}
		\label{wo_enu_9252}
	\end{subfigure}
	\begin{subfigure}[b]{0.48\textwidth}
		\includegraphics[height=0.20in, width=1\textwidth]{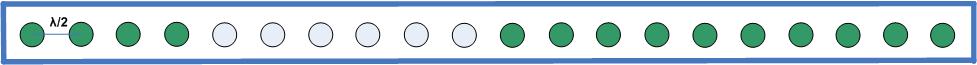}
		\caption{}
		\label{wo_enu_9272}
	\end{subfigure}
	\caption{Example 2 -  (a) Optimum  array TDL implementation scheme (Enumeration)  (b) TDL-SCA    (c) TDL-SDR   (d) DFT-SCA  (e) DFT-SDR (f) Worst case   array (TDL, DFT)}
	\label{op_c_co_c2}
\end{figure}

Sparse array design  utilizes the array aperture and the  additional degrees of freedom, made available by the switching   transceiver chains,  to yield  considerable  SINR improvement. To demonstrate such improvement, the sparse array configuration,  shown in Fig. \ref{wo_enu_ula}, is selected which  occupies the entire array aperture.   It can be observed from the beampattern of this array, shown in Fig. \ref{labelfig5},  that,  in an effort to remove the strong interfering signals, the sparse array failed to provide maximum gain towards the source of interest.  This results in a considerable low output SINR of $0.8$ dB. The actual run times for the SDR-TDL and SDR-DFT  approaches are around 16s and 5s per iteration, respectively (1.4 GHz Quad-Core Intel  i5 processor), whereas it is around 1s for the SCA-TDL and SCA-DFT approaches.  
\subsection{Example 2}
Consider a  wideband source of interest  at $45^0$  and the wideband jammers located at  $35^0$, $55^0$, $60^0$, $145^0$ and $155^0$. A narrowband jammer is located at $135^0$ at an operating frequency of $f_c$, all other parameters are the same as in  Example $1$ except that $P=14$ sensor locations  selected from $N=20$ possible locations and  the wideband source and jammers occupy the entire spectrum from -0.5 to 0.5 cycles/sample.  The SINR of the  optimum array for the TDL implementation scheme  (Fig. \ref{op_enumer_6082}) is $11.32$ dB (found through enumeration). Optimization performed using SCA yields the array in Fig. \ref{op_notap_7752}, with respective  SINR of $11.2$ dB, whereas that performed by SDR yields the array shown in Fig. \ref{op_algo_20272} and corresponding SINR of  $10.9$ dB.   For the dual domain design, the optimum sparse arrays found through DFT-SCA (Fig. \ref{wo_enu_9262})  and through the  DFT-SDR algorithm  (Fig. \ref{wo_enu_9252})  deliver an approximately similar output SINR, around $11.02$ dB. This  is inferior  to the exclusive TDL design.     It is important to  note that the array  configuration resulting in the worst case performance,   shown in  Fig. \ref{wo_enu_9272}, spans  the full  aperture as  the optimum array, yet it offers an output SINR of only $7.35$ dB, underscoring the importance of carefully performing sensor selection for the sparse array design. 
\subsection{Comparison of SDR and SCA under both models }
The design examples discussed thus far   show  amicable  performance of the proposed algorithms under the assumption of the knowledge of full data correlation matrix. The results clearly tie the performance to the location of the sources and their respective powers. However,  evaluating the performance under  matrix completion involves analysis for additional variables, namely, the initial sparse array configuration prior to optimization and the number of snapshots. The performance is, therefore,  dependent on  the observed realization of the received data. In order to have a holistic  assessment of the proposed algorithms,  Monte Carlo simulations   are  generated. We select $P=8$ locations out of  $N=16$ available locations. For  specified DOA of the  desired source,  trials are generated involving six jammers occupying  random locations from $30^0$ to $150^0$.  The SNR of the desired source is $0$ dB, while the  powers  of the jammers are uniformly distributed from $10$ to $20$ dB. The simulation is repeated at $11$ different desired source DOAs, and the average SINR computed. In total, $1500$ experiments are conducted.    For each trial, a random $P$-sparse  array topology serves as an initial  array configuration. This   configuration  could be an optimized configuration from the preceding operating conditions.  The sparse data correlation matrix is estimated based on sensor locations in the initial configuration before performing matrix completion and subsequent optimization process.   The  binary search for the sparsity parameter $\mu$ ranges from $0.01$ to $3$, sparsity threshold $\gamma=10^{-3}$  and $\epsilon=0.05$, relative signal bandwidths and other parameters are the same as given in Example 1.   

Three  benchmarks are established to access the performance of the proposed algorithm under limited snapshots and lack of knowledge of the full data correlation matrix. The first benchmark applies the enumeration technique for MaxSINR design under the assumption that the   data from  all the perspective  sensor locations  is available and  accurate knowledge of the data correlation matrix, i.e., assuming unlimited number of snapshots. This benchmark is referred as ``Knowledge of Full Correlation Matrix-Unlimited Snapshots (KFM-USS)''. Other benchmarks utilize   matrix completion  to recover the missing lags (corresponding to $N-P$ missing sensors).  We refer to these benchmarks as "Matrix Completion-Unlimited Snapshots (MC-USS)" and "Matrix Completion-Limited Snap Shots (MC-LSS)," depending on whether the correlation values are  accurately known through unlimited snapshots or estimated from the limited snapshots. The evaluation  under limited snapshots considers $T=500$.   
\begin{figure}[!t]
	\centering
	\includegraphics[height=2.35in,width=3.68in]{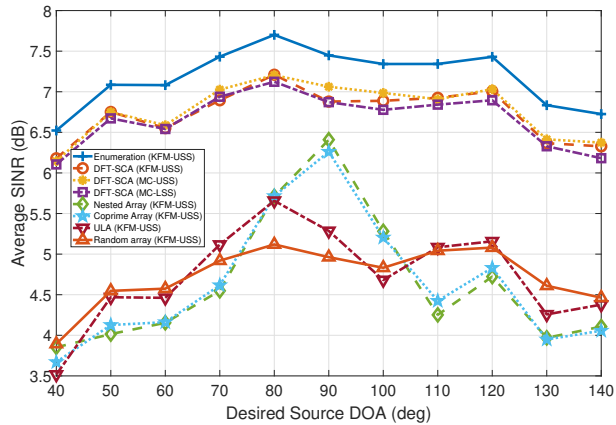}
	\caption{Performance comparisons of SCA under DFT model.}
	\label{labelfig8}
\end{figure}
\begin{figure}[t]
	\centering
	\includegraphics[height=2.35in,width=3.68in]{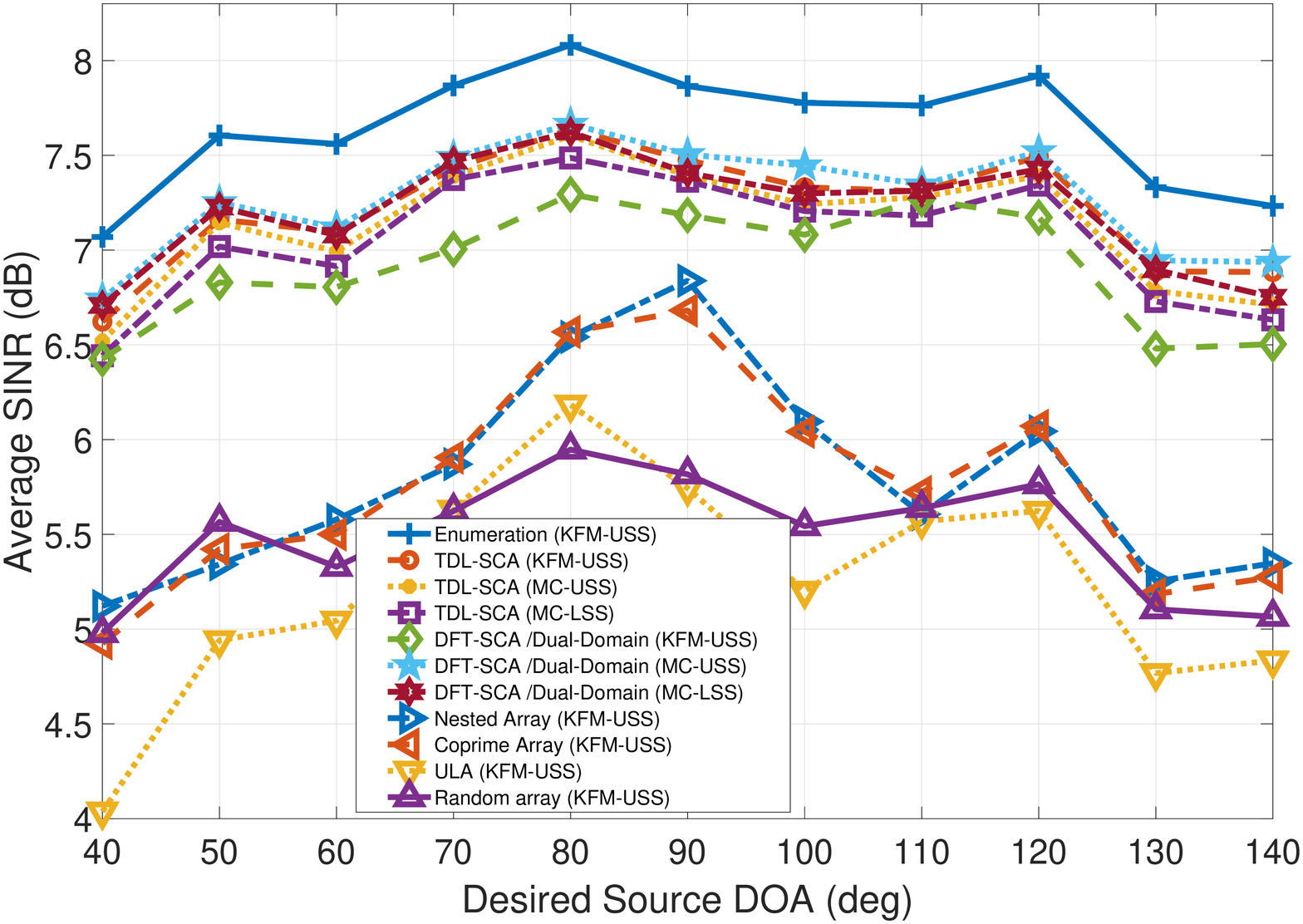}
	\caption{Performance comparisons of SCA under TDL model.}
	\label{labelfig9}
\end{figure}
\begin{figure}[t]
	\centering
	\includegraphics[height=2.35in,width=3.68in]{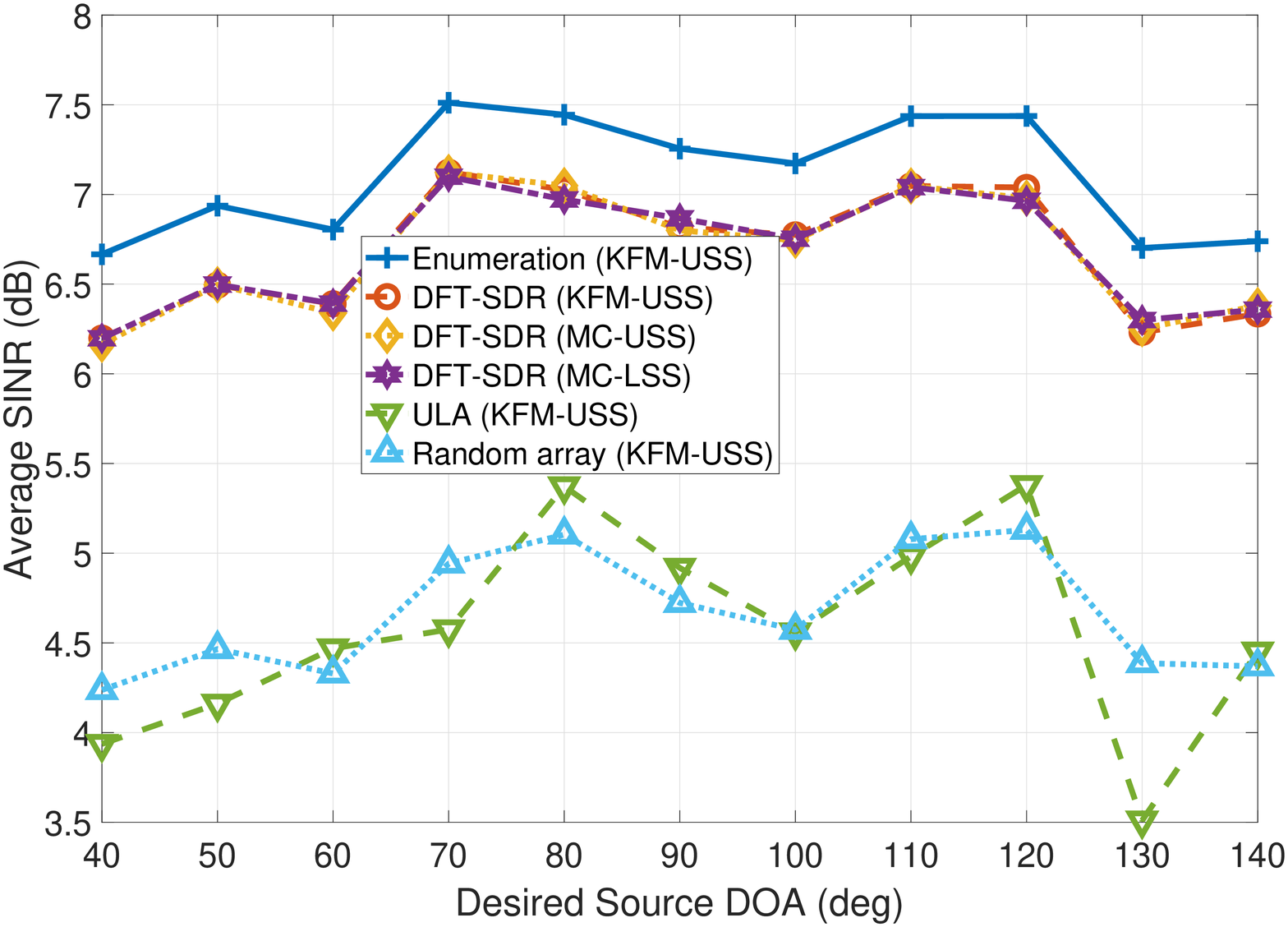}
	\caption{Performance comparisons of SDR under DFT model.}
	\label{labelfig10}
\end{figure}
\begin{figure}[t]
	\centering
	\includegraphics[height=2.35in,width=3.68in]{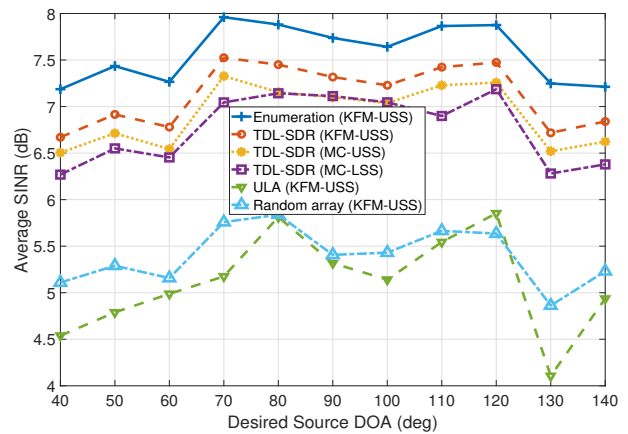}
	\caption{Performance comparisons of SDR under TDL model.}
	\label{labelfig11}
\end{figure}

The performance of the SCA algorithm for the DFT implementation scheme is shown in the Fig. \ref{labelfig8}. The  performance upper bound  is given by the MaxSINR design   evaluated through enumeration (DFT-Enumeration (KFM-USS)). In this case, the  average performance over all the desired source DOAs is  $7.24$ dB.  The proposed  DFT-SCA algorithm under the KFM-USS benchmark  offers an average SINR  of $6.63$ dB. This performance is also comparable to the one achieved through the proposed  matrix completion, as is evident in Fig. \ref{labelfig8}. However,  the DFT-SCA design incorporating the MC-LSS benchmark ($T=500$) has a slight performance trade off of $0.14$ dB w.r.t. the DFT-SCA MC-USS design.  The aforementioned robustness of the MaxSINR design under limited snapshots is partially attributable  to a rather accurate full  matrix estimate achieved by incorporating the apriori knowledge of noise floor. 

The performance of the SCA under the TDL model is evaluated based on the aforementioned benchmarks, as  depicted in Fig. \ref{labelfig9}. The performance  trends are similar, however, the average SINR offered by the TDL implementation scheme is slightly superior to the DFT implementation scheme which is consistent with the literature on wideband beamforming for compact arrays \cite{doi:10.1121/1.413765}. Moreover, it is noted that the DFT-SCA dual domain design achieves comparable performance  to the TDL-SCA under all design benchmarks. This demonstrates the potential of the dual design in achieving an effective  MaxSINR beamformer with reduced complexity.  Figs. \ref{labelfig10} and   \ref{labelfig11} depict the Monte Carlo performance results analyzing the  proposed SDR. It shows that the SDR offers comparable performance to the  SCA technique, but involves  a heavy computational overhead.      It is also clear from the plots that the optimized array configurations offer  a consequential advantage over both the compact ULA and the high resolution structured arrays, such as coprime and nested arrays \cite{5739227,  5456168}, and the randomly selected $P$ sparse array configuration, each employing their respective optimal beamforming weights. The average worst case SINR is, however, reduced significantly to only $1.1$ dB. The performance is also re-evaluated at  varying number of snapshots with  consistent results.  The  performances of the proposed algorithms under MC-LSS inch closer to the MC-USS benchmark with increased data.
\subsection{Practical Considerations for sparse array design} \label{Practical Considerations}
\begin{figure}[!t]
	\centering
	\includegraphics[height=2.05in,width=3.68in]{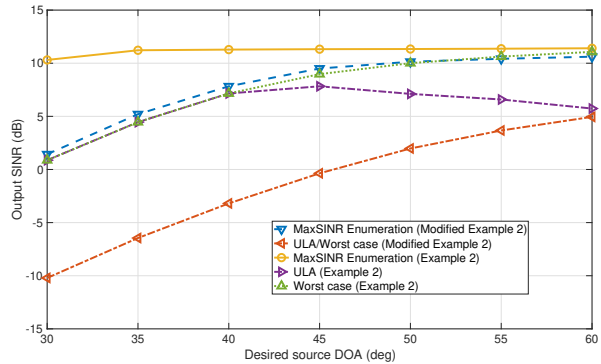}
	\caption{Performance comparisons of the optimum sparse array, the worst performing array and the compact ULA (TDL implementation scheme).}
	\label{labelfig12}
\end{figure}

To assess  the  SINR  advantage of  the optimum sparse array design, we consider the effect of   two important environment dependent parameters, namely, the DOA  of the desired source and the relative locations of the jammers w.r.t. the desired source.    To demonstrate this effect, the desired source DOA in the above examples is  changed  in   steps of $5^0$, with the relative locations of the jammers remaining the same with respect to   the desired source. For example, when the desired source is  at  $50^0$ instead of $45^0$, the corresponding jammer locations shift  by  $5^0$. Figure \ref{labelfig12} compares the performance of the  optimal configuration, the worst performing array, and the compact ULA, with the desired source DOA varying from $30^0$ to $60^0$ under Example 2 and modified Example 2 scenario. In the modified Example 2,  all the parameters are kept the same except the desired source is assumed at $40^0$ instead of $45^0$. It is evident from  Fig. \ref{labelfig12} that  under all  scenarios generated under  modified Example 2, the  compact ULA delivers the worst performance, irrespective of  the  desired source DOA. This is because the  jammers are located closer to the source of interest and the compact ULA  lacks the resolution due to it's  limited array aperture.  On the other hand, for Example 2, the jammers are comparatively widely spaced and as such, the compact ULA  has a satisfactory performance that is close to the optimum sparse arrays, especially when the source DOAs are near the  array broadside. The performance degradation of the ULA near end-fire is due to the increasing  overlap between   the desired signal subspace and the interference plus noise subspace \cite{17564},   therefore lowering SINR performance. In such  scenarios, the sparse array design efficiently utilizes its degrees of freedom to improve SINR by  increasing the separation between the two subspaces. These examples show that the sparse array design is most critical when the underlying jamming environment calls for  additional degrees of freedom to engage the available array aperture more efficiently and to fulfill the  resolution requirements posed by the closer proximity of jammers and  the desired source DOA. 
\section{Conclusion}
This paper considered optimum sparse array design  for maximizing the beamformer output SINR for the case of wideband signal models. Two different implementations, namely the TDL and the DFT implementation schemes, were presented for the optimum sparse array design. The DFT implementation scheme reduces the MaxSINR sparse array design  problem to the lower dimensional space, thereby reducing computational cost. The sparse array configuration optimized in the DFT domain  and later  imported to the TDL implementation scheme is analyzed to alleviate the  computational cost  of the TDL sparse implementation.  It was  shown that the  imported design can possibly yield comparable performance to  the design carried out exclusively through the TDL implementation scheme. For both approaches, we solved the problem using the iterative unit rank promoting SDR algorithm  and a simplified implementation using SCA.  The parameter-free block Toeplitz matrix completion was proposed to realize the data dependent design. It was shown that the SDR and SCA formulation  perform reasonably close to the optimum sparse array design achieved through enumeration under limited data snapshots. The  MaxSINR optimum sparse array yielded considerable performance improvement over suboptimal  sparse arrays and compact ULA for the underlying sensing scenarios. 
\balance
\bibliographystyle{IEEEtran}
\bibliography{references}

\begin{IEEEbiography}[{\includegraphics[width=1in,height=1.25in,clip,keepaspectratio]{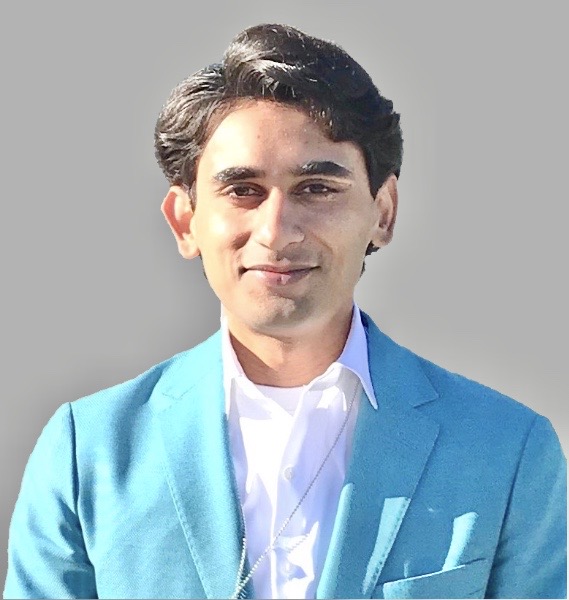}}]{Syed Ali Hamza}
	received the PhD degree in electrical engineering
	from Villanova University, PA, USA in 2020.
	He is currently an Assistant Professor  in the Department of Electrical  Engineering, Widener University, PA, USA. His research interests are statistical and array signal processing,  control systems and distributed signal processing for sensor networks. 
\end{IEEEbiography}
%
\begin{IEEEbiography}[{\includegraphics[width=1in,height=1.25in,clip,keepaspectratio]{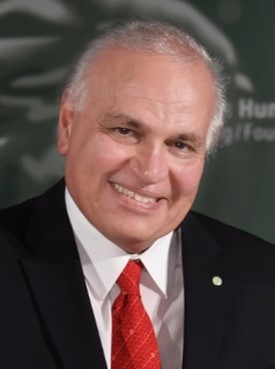}}]{Moeness  Amin}
%
	 is a Fellow of the Institute of Electrical and Electronics Engineers; Fellow of the International Society of Optical Engineering; Fellow of the Institute of Engineering and Technology; and a Fellow of the European Association for Signal Processing. He is the Recipient of: the 2017 Fulbright Distinguished Chair in Advanced Science and Technology; the 2016 Alexander von Humboldt Research Award; the 2016 IET Achievement Medal; Recipient of the 2014 IEEE Signal Processing Society Technical Achievement Award; the 2009 Individual Technical Achievement Award from the European Association for Signal Processing; the 2015 IEEE Aerospace and Electronic Systems Society Warren D White Award for Excellence in Radar Engineering; and the 2010 Chief of Naval Research Challenge Award. Dr. Amin is the Recipient of the IEEE Third Millennium Medal.  He was a Distinguished Lecturer of the IEEE Signal Processing Society, 2003-2004, and is the past Chair of the Electrical Cluster of the Franklin Institute Committee on Science and the Arts.  Dr. Amin has over 750 journal and conference publications in signal processing theory and applications, covering the areas of Wireless Communications, Radar, Sonar, Satellite Navigations, Ultrasound, Healthcare, and RFID.  He has co-authored 21 book chapters and is the Editor of three books titled, Through the Wall Radar Imaging, Compressive Sensing for Urban Radar, Radar for Indoor Monitoring, published by CRC Press in 2011, 2014, 2017, respectively.
\end{IEEEbiography}
\end{document}